\documentclass[%
 reprint,
superscriptaddress,
 amsmath,
 amssymb,
 aps,
 prx,
 showpacs, 
 floatfix,
]{revtex4-2}

\usepackage[margin=1in]{geometry} 
\usepackage{amsmath}       
\usepackage{amssymb}       
\usepackage{braket}        
\usepackage{amsthm}        
\usepackage{hyperref}      
\usepackage[most]{tcolorbox}

\newcommand{\expval}[1]{\left\langle #1 \right\rangle}

\newcommand{\hamil}{\hat{H}}

\usepackage{graphicx}
\usepackage{dcolumn}
\usepackage{bm}
\usepackage{float}
\usepackage{orcidlink}

\usepackage{multirow}
\usepackage{hyperref}
\usepackage[english]{babel}
\hypersetup{
  colorlinks   = true, 
  urlcolor     = blue, 
  linkcolor    = blue, 
  citecolor   = blue 
}
\usepackage{xcolor}
\usepackage{soul}
\sethlcolor{lime}

\begin{document}

\preprint{APS/123-QED}
 
\title{Quantum Birthmarks:\\ Ergodicity Breaking Beyond Scarring}

\author{Anton M. Graf\orcidlink{0000-0002-0573-8907}}
\affiliation{Harvard John A. Paulson School of Engineering and Applied Sciences,
Harvard, Cambridge, Massachusetts 02138, USA}
\affiliation{Department of Physics, Harvard University, Cambridge, Massachusetts 02138, USA}
\affiliation{Department of Chemistry and Chemical Biology, Harvard University, Cambridge,
Massachusetts 02138, USA}

\author{Saul Atwood\orcidlink{0009-0007-0357-3575}}
\affiliation{Department of Physics, Harvard University, Cambridge, Massachusetts 02138, USA}
\affiliation{Harvard College, Harvard University, Cambridge, Massachusetts 02138, USA}

\author{Mingxuan Xiao\orcidlink{0009-0009-4581-8344}}
\affiliation{Department of Physics, Harvard University, Cambridge, Massachusetts 02138, USA}
\affiliation{School of Physics, Peking University, Beijing 100871, China}

\author{Roland Ketzmerick\orcidlink{0000-0002-2518-9343}}
\affiliation{TU Dresden, Institute of Theoretical Physics, 01062 Dresden, Germany}
 
\author{Eric J. Heller\orcidlink{0000-0002-5398-0861}}
\affiliation{Department of Physics, Harvard University, Cambridge, Massachusetts 02138, USA}
\affiliation{Department of Chemistry and Chemical Biology, Harvard University, Cambridge,
Massachusetts 02138, USA}

\author{Joonas~Keski-Rahkonen\orcidlink{0000-0002-7906-4407}}
\affiliation{Computational Physics Laboratory, Tampere University, P.O. Box 600, FI-33014 Tampere, Finland}
\affiliation{Department of Physics, Harvard University, Cambridge, Massachusetts 02138, USA}
\affiliation{Department of Chemistry and Chemical Biology, Harvard University, Cambridge,
Massachusetts 02138, USA}


\date{\today}

\begin{abstract}

A hallmark of classical ergodicity is the complete loss of memory of the initial conditions due to eventual uniform covering of {\it a priori} available phase space. In quantum counterparts of such systems, however, this classical ergodic ideal is fundamentally limited: Here, we introduce the concept of a quantum birthmark, a permanent signature left by the initial state and its early-time evolution in a general quantum system, which gives rise to nonergodic behavior persisting even in the infinite-time limit. We present a birthmark framework outlining a ubiquitous memory effect for an arbitrary, nonstationary state composed of two factors conspiring together: the universal and the revival enhancement. The former sets the minimal amplification carried by the time evolution of a quantum state based on global symmetries, whereas the latter incorporates the further enhancement stemming from the early dynamics, particularly prominent in the presence of recurrences that occur before the Heisenberg time. As a concrete example, we identify quantum birthmarks in the venerable stadium billiard, where they can be significantly enhanced by quantum scars. Finally, we discuss the broader implications of quantum birthmarks, including their role as a natural extension of all types of scarring theories to generic nonstationary quantum systems and prospects for experimental observation. Generally, our work opens an unexplored avenue for understanding the elusive quantum nature of ergodicity.

\end{abstract}

 
\maketitle

\section{Introduction} \label{intro}
\noindent
Thermodynamics rests on an assumption of ergodicity, which in classical systems is associated with chaos~\cite{Gallavotti_book}. Even though chaos plays a crucial role in many scientific fields~\cite{Strogatz_book}, it, along with ergodicity, is a topic fraught with questions and some controversies, often giving rise to misconceptions, particularly in the realm of quantum mechanics (see, e.g., Refs.~\cite{Heller_book, Gutzwiller_book, Haake_book, Tabor_book, Nakamura_book, Casati_book, Stockmann_book}). The correspondence principle states, quite reasonably, that the classical world should start to emerge from the quantum description in the large quantum number, or semiclassical limit, but the role of chaos complicates the situation~\cite{heller_phys.today_46_38_2008, jensen_nature_355_311_1992, jensen_nature_355_591_1992}, as already noted in the early days of quantum mechanics~\cite{stone_phys.today_58_37_2005,einstein_verh.dtsch.phys.ges_19_82_1917}. While there is no quantum version of chaos in the strict sense of classical mechanics~\cite{berry_phys.scripta_40_335_1989, Berry_rspa_413_182_1987}, its presence is nonetheless felt across various circumstances (see, e.g., Ref.~\cite{Stockmann_book}).

\begin{figure}[t!]
    \centering
    \includegraphics[width=1\linewidth]{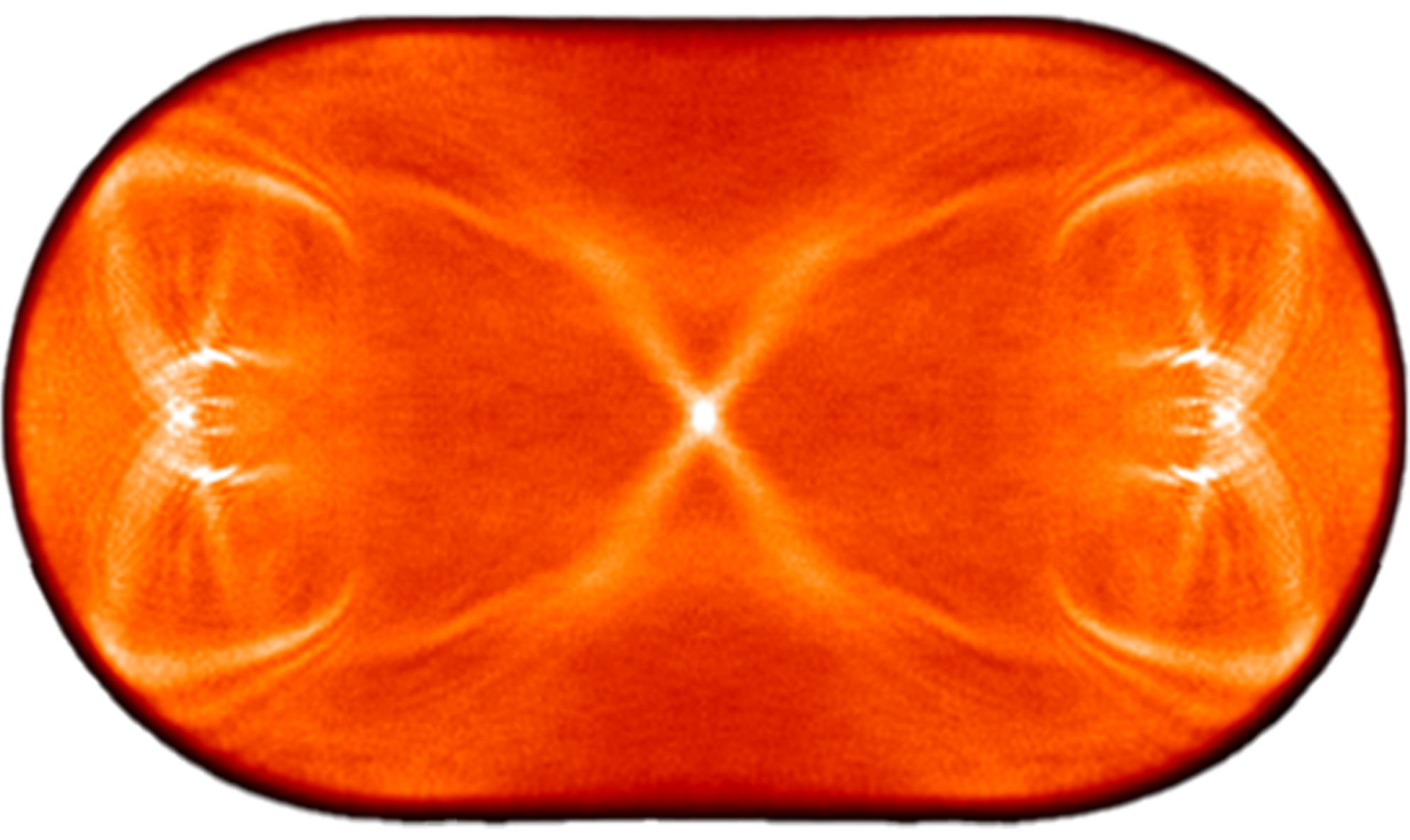}
    \caption{Quantum birthmark: figure displays the long-time average of the probability density for a Gaussian wave packet initialized along a ``bow tie" quantum scar at the center of a stadium billiard (see Sec. III for details). Contrary to the classical expectation of ergodic uniformity, both the initial state and the early-time dynamics induced by the scar are indelibly burned in, manifesting as a quantum birthmark (as explained in Sec. II).\label{fig:example_QB}}
\end{figure}

A widely adopted approach is to search for quantum signatures of chaos either in the statistical analysis of the eigenvalues or in the probability density distribution of eigenstates, both arising from a Hamiltonian whose classical counterpart is chaotic. The former culminates in the famous Bohigas-Giannoni-Schmit (BGS) conjecture~\cite{Bohigas_phys.rev.lett_52_1_1984} rooted in the random matrix theory~\cite{Mehta_book} (RMT): The spectral fluctuations of a generic quantum system are assumed to follow Poisson statistics when the corresponding classical system is integrable, already preceded by the Tabor-Berry conjecture~\cite{berry_proc.r.soc.lond.a_356_375_1997}; whereas they should agree with the Gaussian orthogonal ensemble (GOE) or the Gaussian unitary ensemble (GUE) in the case of a classically chaotic system with or without time-reversal symmetry, respectively. On the other hand, system-specific information is understood within the periodic orbit (PO) theory, especially in the context of the Gutzwiller trace formula~\cite{Gutzwiller_j.math.phys_12_343_1971} connecting the quantum-mechanical spectrum and density of states to the POs in the corresponding classical system.

Concerning eigenstates, there are rigorous quantum-ergodicity theorems~\cite{Shnirelman_Uspekhi.Mat.Nauk_29_181_1974, Colindeverdiere_comm.math.phys_102_497_1985, zelditch_duke.math.j_55_919_1987} stating that the expectation value of an operator converges toward the corresponding classical microcanonical average, which consequently supports the eigenstate thermalization hypothesis (ETH)~\cite{Srednicki_phys.rev.a_50_888_1994, Deutsch_phys.rev.a_43_2046_1991}. Based on the trace formula, wherein each eigenvalue depends on a sum over all POs, it would seem unlikely that a particular PO would stand out in contributing to a particular eigenstate. Moreover, according to the Berry conjecture~\cite{Berry_j.phys.a_10_2083_1977}, a generic eigenstate of a chaotic system can be locally approximated as a random linear combination of plane waves with the local kinetic energy. Nevertheless, not{all eigenstates of a classically chaotic system are doomed to be random and featureless, reflecting the classical ergodicity supplemented with quantum fluctuations.

Crucially, the quantum-ergodicity theorems~\cite{Shnirelman_Uspekhi.Mat.Nauk_29_181_1974, Colindeverdiere_comm.math.phys_102_497_1985, zelditch_duke.math.j_55_919_1987} leave open the possibility of a subset of macroscopically nonergodic eigenstates. A striking visual example is given by a quantum scar (for a review on the topic, see Ref.~\cite{Kaplan_Phys.Today_79_30_2026}). As a consequence of quantum interference~\cite{Heller_phys.rev.lett_53_1515_1984, kaplan_ann.phys_264_171_1998, Kaplan_nonlinearity_12_R1_1999}, the probability density of an eigenstate can be enhanced in the vicinity of a short, moderately unstable PO of the chaotic classical counterpart, thus bearing an imprint of the PO. Significantly, there is no direct analog of scars on the classical side. On the quantum side, they can instead be interpreted as an eigenstate manifestation of short POs correcting the universal RMT eigenvalue statistics~\cite{Berry_proc.r.soc.lond.a_400_229_1985}. In return, scarring of some eigenstates affects the rest of the chaotic (thermal) part of the spectrum of the system in the form of antiscarring~\cite{Antiscarring_1, lu_phys.rev.a_112_043307_2025, Kaplan_phys.rev.e_59_5325_1999}. In fact, quantum scars, living near an unstable PO, are necessarily compensated by corresponding anti-scarred states suppressed along the scar-generating PO to establish together the uniformity of the underlying phase space, as formulated by the stacking theorem~\cite{Antiscarring_1}.

The eigenstate and eigenvalue perspectives both, however, deemphasize the dynamical time evolution of phase-space distributions. Within the time domain, the quantum measures we consider in this work are inspired by classical analogs. In order to assess thermalization or ergodic flow, we start with a localized wave packet in coordinate space, or more generally, in phase space. In classical systems~\cite{Arnold_book}, flux can be slow to leave an original location, yet that location is not favored in any way at long times if the dynamics are ergodic. This notion breaks down dramatically in quantum systems as we highlight here, revealing a profound memory effect -- a \emph{quantum birthmark} (QB) that is permanently imprinted onto the dynamics of the system.

\begin{tcolorbox}[colback=gray!5!white,colframe=gray!75!black,title=Quantum Birthmark, fonttitle=\bfseries]

\textbf{Quantum birthmark} refers to a permanent nonergodic memory of the early dynamics of a nonstationary Hilbert-space state $\vert a \rangle$ and all of its evolutes $\vert a(t) \rangle \equiv \vert \alpha \rangle$ that is reflected in the average occupation probability $\bar{P}_{aa} = \bar{P}_{a\alpha}$ [Eq.~\eqref{Eq:dilution_factor}] over the most nominally matched $\bar{P}_{ab}$ [Eq.~\eqref{Eq:quantum_phase_space_exploration}] given by a ``typical", ergodic state $\vert b \rangle \neq \vert \alpha \rangle$ [Eq.~\eqref{Eq:quantum_ergodic_state}], in a sense that 
\begin{equation*}
    \frac{\bar{P}_{aa}}{\bar{P}_{ab}} \simeq P^{\textrm{UQB}} \cdot P^{\textrm{RQB}} \ge 2.
\end{equation*}
The ratio $\bar{P}_{aa}/\bar{P}_{ab}$ is raised above the ergodic standard (classical expectation) of unity by the universal factor of $P^{\textrm{UQB}} \ge 2$ depending on the global symmetries of the system [Eq.~\eqref{Eq:universal_QB}] and the revival-enhancement factor of $P^{\textrm{RQB}} \ge 1$ encoding the effect of the early dynamics [Eq.~\eqref{Eq:revival_QB}]. 
\end{tcolorbox}

As foreshadowed in Fig.~\ref{fig:example_QB} for a wave packet launched with some mean momentum and position, a memory of its early history is branded into the average occupation probability of the wave packet in the long-time limit. In the far future, it is more likely to return to itself, or any state it evolves into, by a minimum factor of 2 or 3 over most nominally matched wave packets, differing arbitrarily in phase-space origin (different positions and momenta, but the same average energy and energy dispersion). Whether 2 or 3  applies depends on the existence of time-reversal symmetry, and even a higher enhancement applies in the presence of additional symmetries. This enhancement can be boosted further by early-time dynamics, particularly revivals stemming from short POs, or for any reason of slowdown in the exploration of the phase space. Slowdowns are caused by retracing of previously explored zones, a consequence of recurrences. Our notion of QB encapsulates all these aspects into a single entity, as anticipated in the definition above. 

In general, our work contributes to the rapidly growing and far-reaching field of quantum thermalization, shedding light upon the nature of quantum ergodicity. Specifically, we identify a universal constraint governing thermalization and ergodic behavior of quantum systems, ruling out a complete quantum analog of the classical ergodicity hypothesis. This also dampens the emergent idea of Hilbert-space ergodicity~\cite{pilatowsky_phys.rev.lett_131_250401_2023, pilatowsky_phys.rev.x_14_041059_2024}. Our introduction of QBs shifts the focus of quantum ergodicity back to the time domain, drawing more of a parallel with its classical counterpart, which is almost always cast in terms of time evolution. 

Moreover, although quantum scars are intimately tied to the time domain, as made clear by the first proof of their origin \cite{Heller_phys.rev.lett_53_1515_1984}, they remain as structures seen in the energy domain. Our QB theory offers a more general framework incorporating scar phenomenology into a larger context beyond scarring. In fact, the important extensions of the original scar theory \cite{Heller_phys.rev.lett_53_1515_1984,kaplan_ann.phys_264_171_1998,Kaplan_nonlinearity_12_R1_1999}, including variational scarring~\cite{keski-rahkonen_j.phys.conden.matter_31_105301_2019, keski-rahkonen_phys.rev.b_97_094204_2017, keski-rahkonen_phys.rev.lett_123_214101_2019, selinummi_phys.rev.b_110_235420_2024, keskirahkonen_phys.rev.e_112_L012201_2025, chalangari_phys.rev.b_112_115137_2025, luukko_sci.rep_6_37656_2016} and many-body scarring~\cite{serbyn_nat.phys_17_675_2021, chandran_annu.rev.condens.Mmatter.phys_14_443_2023, evrard_phys.rev.lett_132_020401_2024}  phenomena, can also become part of the larger concept of QBs. We elaborate on this point below while presenting the emergence of QBs and analyzing their broader and experimental significance.

The manuscript is organized as follows: In Sec.~\ref{memory}, we develop and establish the concept of QBs in detail. Section~\ref{models} demonstrates the concept of universal and also revival-enhanced QBs, employing the prototypical model of the chaotic Bunimovich stadium. In Sec.~\ref{Sec:discussion}, we map out connections between QBs and other areas of the field, as well as outline potential directions for future research. Finally, we conclude with a brief summary in Sec.~\ref{Sec:conclusion}.

\section{Quantum memory} \label{memory}

We start by considering two generic nonstationary states $\vert a \rangle$ and $\vert b \rangle$ in an $N$-dimensional Hilbert space spanned by the basis of eigenstates $\vert E_n \rangle$ ($n = 1, \dots, N$) of the given Hamiltonian $\mathcal{H}$ with the corresponding energies $E_n$. For simplicity, we can assume that the spectrum of the Hamiltonian is nondegenerate.  Then, we can decompose the state $\vert a \rangle$ as
\begin{equation}\label{Eq:expansion_of_the_initial_state}
    \vert a \rangle = \sum_n a_n\vert E_n \rangle
    \quad \textrm{with} \quad
    a_n = \langle E_n \vert a \rangle,
\end{equation}
and a similar expansion exists for the state $\vert b \rangle$. Moreover, let us denote the corresponding time-evolved states as
\begin{equation}
   \vert \alpha \rangle = \mathcal{U}(t) \vert a \rangle 
   \quad \textrm{and} \quad
   \vert \beta \rangle = \mathcal{U}(t) \vert b \rangle, 
\end{equation}
where $\mathcal{U}(t)$ is the time-evolution operator arising from the Hamiltonian $\mathcal{H}$~\footnote{Here, we also allow backward time evolution, i.e., $\vert \alpha \rangle = \mathcal{U}(-t) \vert a \rangle$. Throughout the text, we implicitly assume $t \ge 0$, except for a brief discussion in Sec.~\ref{models}. Nonetheless, our results remain valid if the time order is reversed or extended to the limit $t \rightarrow -\infty$.}.

Motivated by Refs.~\cite{Smith_phys.rev.e_80_035205_2009, Smith_phys.rev.e_82_016214_2010}, we begin by preemptively defining a key time-domain observable, which we refer to as the autocorrelation function or generalized fidelity. This quantity captures both transition and survival amplitudes and is useful for probing correlations between different initial and final states under unitary time evolution. For the chosen states $\vert a \rangle$ and $\vert b \rangle$, it is defined, in a similar manner to Refs.~\cite{Smith_phys.rev.e_80_035205_2009, Smith_phys.rev.e_82_016214_2010}, as
\begin{equation}\label{Eq:generalized fidelity}
\begin{split}
    P_{a\beta}(t) &= \frac{1}{2}\Big( \vert \langle a \vert  \beta \rangle\vert^2 + \langle a \vert \alpha\rangle \langle \beta \vert b \rangle \Big)\\
    &=\frac{1}{2}\Big( \vert \langle a \vert \mathcal{U}(t) \vert b \rangle\vert^2 + \langle a \vert \mathcal{U}(t) \vert a \rangle \langle b \vert \mathcal{U}^{\dagger}(t) \vert b \rangle \Big),
\end{split}    
\end{equation}
which can be interpreted as a measure for how much a test state $\vert a \rangle$ is transferred to a probe state $\vert \beta \rangle$ by the quantum evolution. The primary motivation for introducing the generalized fidelity in this form is its ability to encode the relative time evolution of the states $\vert a \rangle$ and $\vert b \rangle$, which is subsequently exploited to define the revival part $P^{\textrm{RQB}}$ of a QB carrying on signatures of the early-time dynamics presented later in Eq.~\eqref{Eq:revival_QB}.

As illustrated in Fig.~\ref{fig:generalized_fidelities}, the generalized fidelity distinguishes between  regular and chaotic quantum dynamics, particularly in the long-time limit. Whereas the evolution of a quantum-chaotic system can, at least, lead to partial overlap of the probe state $\vert \beta \rangle$ with the test state $\vert a \rangle$, yielding $P_{a \beta}(t) \neq 0$; this possibility can be strongly suppressed, i.e., $P_{a \beta}(t) \sim 0$ in a regular system. In particular, we later employ the generalized fidelity to embed early-time dynamical features into the otherwise universal predictions of RMT [see Eq.~\eqref{Eq:revival_QB}].

In the special case $\vert a\rangle=\vert b\rangle$, the generalized fidelity reduces to the conventional autocorrelation function.
\begin{equation}\label{Eq:fidelity}
    P_{a \alpha}(t) = \vert \langle a \vert \alpha \rangle \vert^2 = \vert \langle a \vert \mathcal{U}(t) \vert a \rangle \vert^2.
\end{equation}
In systems with chaotic dynamics, the fidelity is often exponentially damped due to the instability of trajectories, as bounded by ${\rm max}[P_{a \alpha}(t)] \sim e^{-\lambda t}$, where $\lambda$ is the Lyapunov exponent~\cite{Heller_book}. However, in quantum systems, the discreteness of energy levels ensures that recurrences continue indefinitely, albeit at diminished amplitudes.

\begin{figure}[ht]
    \centering
    \includegraphics[width=1.0\linewidth]{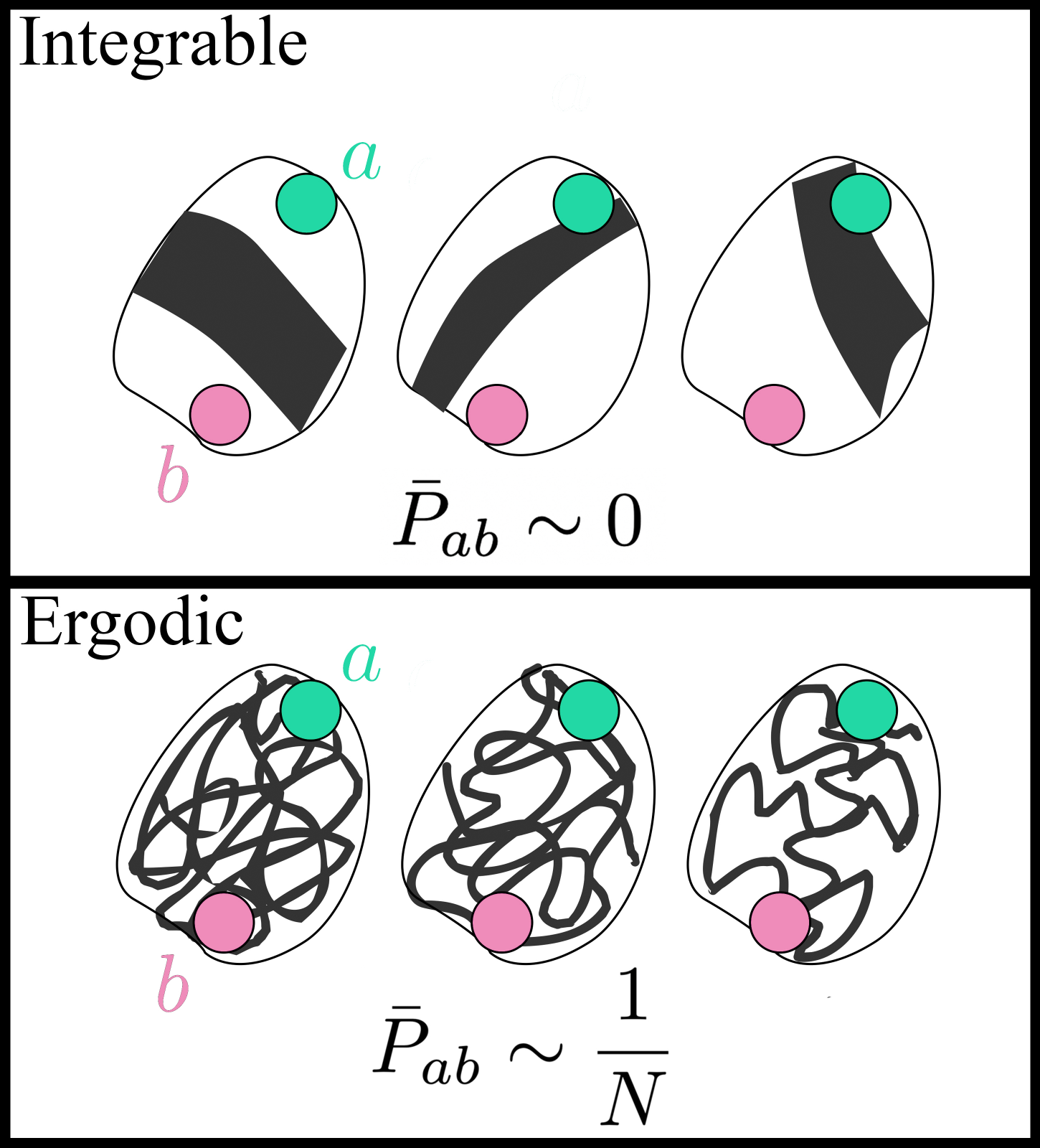}
    \caption{Average occupation probability [see Eq.~\eqref{Eq:quantum_phase_space_exploration}]: The localized and ``dynamically distinguishable" states $\vert a \rangle$ and $\vert b \rangle$ are depicted schematically as circles, alongside representative spatial profiles of typical eigenstates of the considered system. At the top, corresponding to an integrable system, each eigenstate overlaps with either $\vert a \rangle$, or $\vert b \rangle$, or neither -- importantly not both, in the cases shown. If  $\vert b \rangle$ had been in the zone reached by $\vert a \rangle$, the overlap would be large. At the bottom, representing an ``ergodic'' system, each eigenstate overlaps with both $\vert a \rangle$ and $\vert b \rangle$. This highlights a fundamental distinction between regular and chaotic quantum dynamics. In the long-time limit ($t \to \infty$), the average occupation probabilities behave differently in the two cases. For integrable systems, one finds $\bar P_{ab} \to 0$ (or more generally, $\bar P_{ab} \sim m/N$, with $m$ of order $N$), provided that the probe state lies within the manifold accessed from $\lvert a\rangle$. By contrast, for chaotic systems, the long-time average approaches the ergodic value $\bar P_{ab} \to 1/N$.
    }
\label{fig:generalized_fidelities}
\end{figure}

We turn next to the time-averaged probability $\bar{P}_{ab}$ of occupying the state $\vert b \rangle$ if initially the system is in the state $\vert a \rangle$. This is given by the inner product
of their respective spectra or eigenstate probability distributions~\cite{Heller_book}: 
\begin{equation}\label{Eq:quantum_phase_space_exploration}
   \bar{P}_{ab} = \lim_{T \rightarrow \infty} \frac{1}{T} \int_0^T \vert \langle b\vert \mathcal{U}(t)\vert a \rangle \vert^2 \, dt = 
   \sum\limits_n p_n^{a} p_n^{b},
\end{equation}
with the individual probability distribution $p_n^{a} = \vert a_n \vert^2$ and $p_n^{b} = \vert b_n \vert^2$ of being in the eigenstate $\vert E_n \rangle$ for the nonstationary state $\vert a \rangle$ and $\vert b \rangle$, respectively~\cite{Heller_book}. Notably, even though the definition of the time-averaged quantity $\bar{P}_{ab}$ is different from its time-dependent counterpart $P_{\alpha\beta}(t)$, they are in fact consistent with each other in the sense of $\bar{P}_{ab} = \overline{P_{\alpha \beta}(t)}$~\footnote{Let us use the eigenstate expansion for the both terms of the generalized fidelity in Eq.~\eqref{Eq:generalized fidelity}, and then time averaged these terms. Since it is assumed that there are no degeneracies in the spectrum and the time evolution operator is $\mathcal{U}(t) = \exp(i \frac{\mathcal{H}}{\hbar}t)$, the Riemann-Lesbegue lemma yields us that $\overline{\vert \langle a \vert \mathcal{U}(t)\vert b \rangle \vert^2} = \sum_n p_n^a p_n^b$ and $\overline{\langle a \vert \mathcal{U}(t) \vert a \rangle \langle b \vert \mathcal{U}^{\dagger}(t) \vert b \rangle} = \sum_n p_n^a p_n^b$. Therefore, we see $\overline{P_{\alpha \beta}(t)} = \sum_n p_n^ap_n^b$ that is the same as the time-averaged probability $\bar{P}_{ab}$ defined in Eq.~\eqref{Eq:fidelity}.}.

Naturally, this idea translates into the language of density matrices that determine the time-averaged probability for the density matrix $\rho_b = \vert b \rangle \langle b \vert$ when the initial state is described by the density matrix $\rho_a = \vert a \rangle \langle  a \vert$ as
\begin{equation}\label{Eq:quantum_phase_space_exploration_density_matrix}
\begin{split}
   \bar{P}_{ab} &= \lim_{T \rightarrow \infty} \frac{1}{T} \int_0^T  \text{Tr} \left[ \rho_b \mathcal{U}(t)\rho_a \mathcal{U}^{\dagger}(t) \right] \, dt =\sum\limits_n p_n^{a} p_n^{b},
\end{split}
\end{equation}
with the updated notation for the individual probabilities $p_n^{a} = \vert \langle n \vert \rho_a \vert n \rangle \vert$ and $p_n^{b} = \vert \langle n \vert \rho_b \vert n \rangle \vert$. In the special case of the identical states, we have
\begin{equation}\label{Eq:dilution_factor}
\begin{split}
    \bar{P}_{aa} &=  \lim_{T \rightarrow \infty} \frac{1}{T} \int_0^T P_{a\alpha}(t) \, dt\\
    &=\lim_{T \rightarrow \infty} \frac{1}{T} \int_0^T \vert \langle a \vert \mathcal{U}(t)\vert a \rangle \vert^2 \, dt = \sum_n (p_n^a)^2
\end{split}    
\end{equation}
which is aptly referred to as the dilution factor and coincides with the standard definition of the inverse participation ratio introduced later in Eq.~\eqref{Eq:participation_ratio}. As in the case of Eq.~\eqref{Eq:quantum_phase_space_exploration}, when identifying the time average in Eq.~\eqref{Eq:quantum_phase_space_exploration_density_matrix}, we have assumed nondegeneracy of the energy spectrum.

\begin{figure*}[t!]
    \centering
    \includegraphics[width=\textwidth]{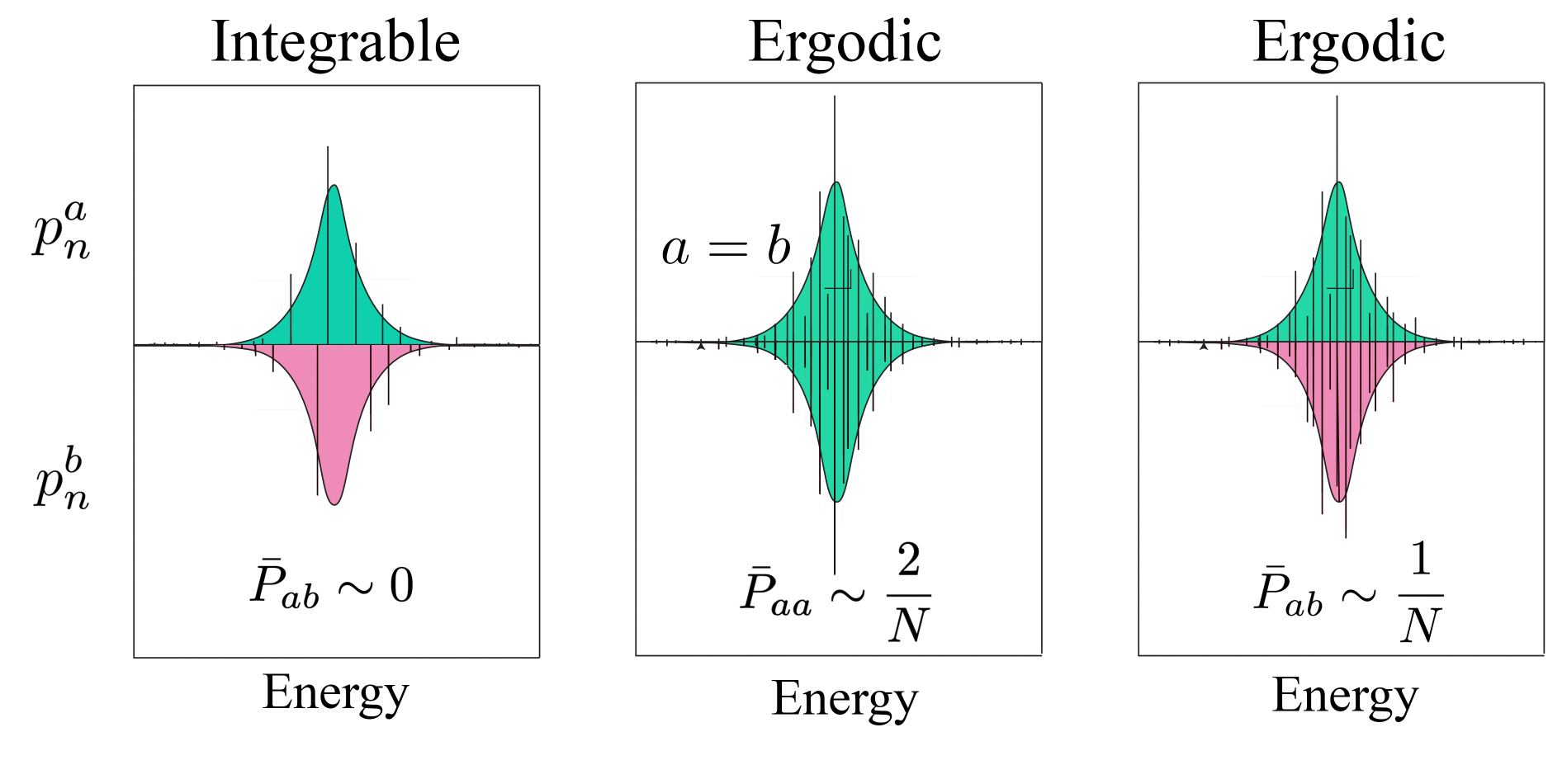}
    \caption{Spectral decomposition and overlap: In the left panel corresponding to integrable behavior, the states $\vert a \rangle$ and $\vert b \rangle$ have mutually exclusive occupation of the eigenstates, and vanishing dynamical access of one starting from the other. However, in the chaotic system shown in the middle, the spectral intensities are fluctuating according to a $\chi^2$ distribution of 2 degrees of freedom (the square of Gaussian distributed amplitudes). Since here the large fluctuations are always squared, the sum $\sum_n (p_n^a)^2$ rises to $2/N$, as discussed in Sec.~\ref{Sec:UQP}. On the right, the states $\vert a \rangle$ and $\vert b \rangle$ share the same envelopes, but differ in their overlaps with the eigenstates. In this case, it is random as to whether large probabilities in $p_n^a$ correspond to large probabilities in $p_n^b$, yielding the sum $\sum_n p_n^a p_n^b$ matching with the ergodic expectation of $1/N$.}
    \label{Fig:Pabspectra}
\end{figure*}

Both forms [Eqs.~\ref{Eq:quantum_phase_space_exploration} and \ref{Eq:quantum_phase_space_exploration_density_matrix}] of this measure of ergodicity are commutative with respect to the chosen states, i.e., $\bar{P}_{ab} = \bar{P}_{b a}$.  Figure~\ref{Fig:Pabspectra} illustrates the motive behind this spectral decomposition of the averaged occupation probability for an integrable and quantum-chaotic system. Regarding the phase-space flow, there is no distinction between the state $\vert a \rangle$ and any of its time-evolved counterparts  $\vert \alpha \rangle$, implying that
\begin{equation}\label{Eq:time_invariance_of_quantum_measure}
    \bar{P}_{aa} = \bar{P}_{a \alpha} = \bar{P}_{\alpha a} = \bar{P}_{\alpha \alpha}.
\end{equation}
Therefore, when considering two arbitrary states $\vert a \rangle$ and $\vert b \rangle$, we normally assume them to be dynamically distinguishable, that is, they are not related by the time evolution $\mathcal{U}(t)$. 

The distinguishable states may, of course, have nonzero overlap, and be similar in their energy envelope and dispersion. Therefore, to give further significance to the measure $\bar{P}_{ab}$, we must develop reasonable prior constraints omitting cases of trivial nonergodicity, such as testing flow between states $\vert a \rangle$ and $\vert b \rangle$ that have nonoverlapping energy envelopes. We normally include that envelope as an \textit{a priori} constraint (partial overlap can be handled as presented in Ref.~\cite{heller_phys.rev.a_35_1360_1987}) and thus focus on questions of ergodicity that remain interesting apart from things we know very early, such as mean energy and energy dispersion. All states that concur in these measures are called ``{\it a priori} matched''. 

Classical ergodicity can be defined as the time-averaged probability distribution becoming uniform across the phase-space energy hypersurface, when energy is the only relevant constraint. However, we cannot similarly specify the energy in the quantum case, and instead we specify {\it a priori} matched states as those sharing the same average energy and its dispersion~\footnote{This determination of {\it a priori} matched
quantum states is formally expressed as the infinite time average of a trajectory becoming, with infinitesimal coarse graining, a uniform phase distribution of
\begin{equation}
    \rho_\infty({\bf p,q}) = \frac{\delta (E-H({\bf p,q}))}{{\rm Tr}(\delta (E-H({\bf p,q})))},
\end{equation} where $H$ is the classical Hamiltonian and $E$ the energy of the particle exploring the phase space.}. Within the Hilbert space of $N$ such states, the statement that all states are equally accessed in the quantum-ergodic limit is taken as
\begin{equation}\label{Eq:quantum_ergodic_state}
    p_n^{a \,\textrm{or} \, b} = \frac{1}{N} \quad \forall\, n \, \Rightarrow \, \bar{P}_{ab} = \frac{1}{N}.
\end{equation}
Starting in any one of the $N$ states, complete ergodicity requires it to be equally likely to be found in any other {\it a priori} matched final state in the sense of an infinite time average. This criterion is commonly regarded as the gold standard for quantum ergodicity (see, e.g., Refs.~\cite{pilatowsky_phys.rev.lett_131_250401_2023, pilatowsky_phys.rev.x_14_041059_2024}).

A key message here is that the classical idea of ergodicity and its direct quantum version stated in Eq.~\eqref{Eq:quantum_ergodic_state} require a modification to account for both short-time and long-time dynamical events which are never forgotten in quantum systems. We mean this in the sense of an enhancement within the time-averaged occupation probability ratio, composed as~\cite{Smith_phys.rev.e_80_035205_2009, Smith_phys.rev.e_82_016214_2010}
\begin{equation} \label{Eq:bootstrap}
    \frac{\bar{P}_{aa}}{\bar{P}_{ab}} \approx P^{\textrm{UQB}} \cdot P^{\textrm{RQB}} \geq 2 ,
\end{equation} 
where  $\vert a\rangle$ and $\vert b\rangle$ are members of the {\it a priori} matching set. This value is greater than 1 corresponding to the classical ergodicity expectation. We will show that $P^{\textrm{UQB}} \ge 2$, a universal quantum-birthmark (UQB) factor, and $P^{\textrm{RQB}} \ge 1$ a revival-enhanced quantum-birthmark (RQB) factor, respectively.  These are defined in context, further defined in Eqs.~\ref{Eq:universal_QB} and \ref{Eq:revival_QB}, below. This decomposition already implies the breakdown of ``perfect" quantum ergodicity of $\bar{P}_{aa} = \bar{P}_{ab}$, as well as a breakdown of the spirit of the strong ETH. 

According to Eq.~\eqref{Eq:bootstrap}, every initial quantum state $\vert a \rangle$, along with the states $\vert \alpha \rangle$ it evolves into, experiences a universal enhancement factor of $P^{\textrm{UQB}}$ arising even if the spectrum has qualities expected of a random matrix (see, e.g., Ref.~\cite{Kaplan_phys.rev.e_59_5325_1999}), whereas the revival factor $P^{\textrm{RQB}}$ takes into account short-time dynamics, correcting the RMT behavior for recurrences which might be due to a PO for example (see, e.g., Refs.~\cite{Smith_phys.rev.e_80_035205_2009, Smith_phys.rev.e_82_016214_2010}). We explore these factors in more depth below.

We refer to this permanent enhancement of an arbitrary state and its temporal track as QB (see Sec.~\ref{intro}), composed of a universal factor ($\sim P_{ab}^{\textrm{UQB}}$) and a revival-enhanced  factor ($\sim P_{ab}^{\textrm{RQB}}$) . A QB is an inevitable consequence of  quantum evolution following a launch of a nonstationary state. The existence of such a UQB is not intrinsic to the system, and it is independent of POs or scarring. On the other hand, if an initial state is aligned with a PO, a revival-enhanced birthmark materializes along the orbit. As encoded in RQB, a partial revival implies retracing of previously visited parts of Hilbert space, which could enhance or detract from those regions, depending on the returning phase at the given time. Quantum scars are of this nature, and thus the concept of an RQB becomes a natural generalization of the phenomenon. Nevertheless, recurrences and stronger QBs may even take place in the absence of nearby POs.

To develop the QB theory here, we first address the classical phase-space analog that underpins our measure $\bar{P}_{ab}$, leading to the maximum rate principle. We then move on presenting a more rigorous framework for the universal $P^{\textrm{UQB}}$ and lastly for the revival-enhanced component $P^{\textrm{RQB}}$ of a QB.

\subsection{Phase-space exploration}

Classical ergodicity implies that all eligible regions of phase space are eventually explored uniformly by a chaotic trajectory over a hypersurface defined by stipulated constraints, including fixed energy and angular momentum. In contrast, quantum systems with sharply defined energy (i.e., the eigenstates of the Hamiltonian) have no dynamics, despite the fact that classically, sharply defined energy does not preclude rich dynamical behavior. To meaningfully characterize quantum dynamics, we must therefore turn to nonstationary states with finite energy spreads. Indeed, well-defined classical analogs exist for such states, corresponding to ensembles or distributions that span a range of energy shells (see, for instance, Ref.~\cite{Heller_book}).

A nonstationary quantum state is characterized by an energy expectation value and dispersion, both of which can be inferred from short-time dynamics. The same applies to any final state employed  to probe or quantify phase-space flow. Crucially, pairs of nonstationary states with disjoint energy support cannot exhibit a dynamical flow between them over time, rendering them uninteresting as a pair for  analysis. This applies in both the quantum and classical domains. The classical and quantum descriptions can be brought into close correspondence by employing wave packets as initial states and representing them with their associated Wigner distributions in phase space. Then, in either case, when states exhibit only partial energy overlap, appropriate weighting or biasing schemes must be introduced to properly bias the measure, ergo making the outcome informative~\cite{heller_phys.rev.a_35_1360_1987}. 

Defining the phase-space exploration rate of a nonstationary quantum state, we consider a distribution $\rho$ over a $D$-dimensional phase space $(\boldsymbol{q}, \boldsymbol{p})$, which is normalized as
\begin{equation}
\text{Tr} (\rho) = \int \int \rho (\boldsymbol{q},\boldsymbol{p}) d\boldsymbol{p} d\boldsymbol{q} = 1.
\end{equation}
The phase-space distribution $\rho (\boldsymbol{q},\boldsymbol{p})$ can be purely classical or a Wigner transform of a quantum wavefunction or density.
In the special case where the density is uniform across the phase space, formally $\rho = V^{-1}$, we immediately find that 
\begin{equation*}
    \text{Tr}\left(\rho^2 \right) = \int \int\rho^2(\boldsymbol{q},\boldsymbol{p}) d\boldsymbol{p} d\boldsymbol{q} = \frac{1}{V}
\end{equation*}
which establishes a familiar metric for phase-space exploration, aligning with the dilution factor $\bar{P}_{aa}$ in Eq.~\eqref{Eq:dilution_factor}. The quantity $\textrm{Tr}(\rho^2)$ serves as a reasonable definition of the associated phase-space volume for smoothly varying distributions. If the density $\rho$ corresponds to a pure state, the volume occupied is one Planck cell of $h^{D}$, or equivalently
\begin{equation}
\int \int \rho^2 (\boldsymbol{q}, \boldsymbol{p}) d\boldsymbol{p} d\boldsymbol{q} = \frac{1}{h^D}.
\end{equation}

We turn our attention to  the more interesting case of a nonstationary distribution $\rho(\boldsymbol{q},\boldsymbol{p},t)$ evolving under the influence of the Hamiltonian $\mathcal{H}$.  Estimating the rate at which the distribution $\rho(t)$ explores new regions of phase space requires a measure that captures the cumulative extent of its prior visitation. An appealing choice is the time-averaged distribution
\begin{equation}
\bar{\rho} (\boldsymbol{q}, \boldsymbol{p}, t) = \frac{1}{t} \int_0^t \rho (\boldsymbol{q}, \boldsymbol{p}, \tau) d \tau,
\end{equation}
which is normalized, and yields a measure for the number $\mathcal{N}_t$ of accessed phase-space cells by the time $t$ as
\begin{equation}
\frac{1}{\mathcal{N}_t} = h^D\text{Tr} [ \bar{\rho}^2 ] \leq 1.
\end{equation}
Combining the equations above supplemented with the fact $\text{Tr} [ \rho (\tau) \rho (\tau') ] = \text{Tr} [ \rho(0) \rho (\tau'-\tau) ]$ invoking the time-translation symmetry~\footnote{This is justified because we here focus on a time-independent Hamiltonian of a closed system.}, we can write~\cite{heller_phys.rev.a_35_1360_1987} 
\begin{equation}\label{Eq:accessed_phase_space_cells}
\frac{1}{\mathcal{N}_t}= h^D \text{Tr} [ \bar{\rho}^2 ] = \frac{2}{t} \int_0^t \left[ 1 - \frac{\tau}{t} \right] P(\tau) d \tau,
\end{equation}
where we have defined the survival probability as
\begin{equation}
    P(\tau)= h^D \text{Tr} [ \rho(0) \rho (\tau) ].
\end{equation}
The survival probability $P(\tau)$ plays a similar role to the fidelity $P_{a\alpha}(t)$ previously defined in Eq.~\eqref{Eq:fidelity}. Moreover, the time-averaged density $\bar{\rho}(p,q,t)$ is nonzero only in regions that have been accessed by the distribution $\rho (p,q,t)$, but in the manner of ``feathering out" rather than cutting off abruptly. As a result, the distinction between covered and uncovered regions is smoothed, and the number $\mathcal{N}_t$ of explored phase-space cells then extends also to account for partial visitations.

With the number of phase-space cells determined according to Eq.~\eqref{Eq:accessed_phase_space_cells}, the associated rate $\mathcal{R}_t$ of phase-space exploration is defined~\cite{heller_phys.rev.a_35_1360_1987} as
\begin{equation}\label{eq:phase_space_rate}
\mathcal{R}_t = \frac{d {\mathcal{N}_t}}{d t}= \frac{\frac{1}{2} \int_{0}^{t} \left( 1 - \frac{2\tau}{t} \right) P(\tau) d\tau }{\left[ \int_{0}^{t} \left( 1 - \frac{\tau}{t} \right) P(\tau) d\tau \right]^2}.
\end{equation}
Notably, the rate $\mathcal{R}_t$ is not guaranteed to be strictly positive. Recurrences or salvaging of previously explored regions of phase space leads to an uneven distribution $\bar{\rho}$, which temporarily causes the state count $\mathcal{N}_t$ to diminish. On the other hand, for large enough $t$ but before any recurrences ($\sim T_{\text{min}}$), the exploration rate above approaches the initial-decay steady-state value of
\begin{equation}\label{eq:phase_space_rate_maximum}
\mathcal{R}_{\textrm{max}} = \frac{1}{2 \int_{0}^{T_{\textrm{min}}} P(\tau) d\tau}.
\end{equation} 
This maximal rate $\mathcal{R}_{\textrm{max}}$ is determined by the time $T_{\textrm{min}}$, which can generally be identified with the onset of the first (partial) recurrence, typically governed by the shortest PO. Particularly, the notion of the quantum rate of phase-space exploration~\cite{heller_phys.rev.a_35_1360_1987} provides a dynamically grounded extension of the conventional participation ratio approach (see, e.g., Ref.~\cite{bies_phys.rev.e_63_066214_2001}):
\begin{equation}\label{Eq:participation_ratio}
    \Gamma^{-1} = \frac{1}{\sum_n {p_n^{a}p_n^{a}}} = \frac{1}{\sum_n \vert a_n \vert^{4}}.
\end{equation}

The initial decay  of the survival probability $P(\tau)$ sets a rate of exploration $\mathcal{R}_{\textrm{max}}$, which can decrease only with future recurrences. The maximum rate principle sets a standard: Partial revivals lead to the reaccessing of regions in phase space that have already been explored, slowing the rate of exploration of new phase space below the maximum rate. Importantly, the rate of travel remains fixed as the distribution $\rho(\boldsymbol{q}, \boldsymbol{p}, t)$ sweeps through the phase space, but the rate $\mathcal{R}_t$ of exploration of virgin, previously unexplored phase-space territories declines after a revival. Revivals imply that a phase-space domain already visited is to be revisited which is a less efficient process. This behavior is true classically as well, but classical Hamiltonian systems have all the time in the world to scout through the phase space.

Importantly, quantum systems exhibit a kind of ``Cinderella effect'': the discovery of new regions of phase space effectively ceases approximately at the Heisenberg time, \( t_H \simeq \hbar / \delta E \), where $\delta E$ denotes the mean level spacing. At this timescale, adjacent energy levels begin to dephase from one another, and further exploration can cover only old ground. This concept is aided by the level repulsion characteristic of chaotic systems, discouraging eigenvalue pairs much closer than the average level spacing, as present in the BGS conjecture~\cite{Bohigas_phys.rev.lett_52_1_1984}. 

The {\it a priori} available phase space is determined by the density of states times the energy width of the energy envelope. That envelope is, so to speak, set by the user, and is set by the initial-decay time of the chosen nonstationary state. Furthermore, the maximum rate principle dictates that the whole phase space cannot be fully discovered if the rate of exploration of new phase space has been slowed earlier by any recurrence or revivals. Yet, the quantum-ergodicity game is even more rigged from the beginning: As demonstrated below, a nonstationary state retains a permanent memory of its past through the presence of a UQB, which signals a breakdown of ergodicity and is further amplified by the possible emergence of an RQB.

\subsection{Universal quantum birthmarks}\label{Sec:UQP}

Even in the absence of early-time revivals, or $P^{\textrm{RQB}} = 1$, it is very interesting that the measure in Eq.~\eqref{Eq:quantum_phase_space_exploration} predicts a rather large memory effect, purely due to quantum interference. Based on Eq.~\eqref{Eq:dilution_factor}, the probability that the
system has returned to the initial state $\vert a \rangle$ is
\begin{equation}\label{Eq:dilation_and_PR}
    \bar{P}_{aa} = \lim_{T \rightarrow \infty} \frac{1}{T} \int_0^T \vert \langle a \vert \mathcal{U}(t) \vert a \rangle \vert^2 \, dt = \sum_n \vert a_n \vert^4.
\end{equation}
In other words, the average occupation probability $\bar{P}_{aa}$ reduces to the participation ratio in Eq.~\eqref{Eq:participation_ratio}. Interestingly, under the unitary evolution, there is no difference between regular and chaotic dynamics, or in the terms of the BGS conjecture~\cite{Bohigas_phys.rev.lett_52_1_1984} as there is no energy spectrum dependence. Rather the dilution $\bar{P}_{aa}$ depends only on the initial state, i.e., on the coefficients $a_n$, in sharp contrast to the classical dynamics. 

It can be seen that there is a quantum bound
\begin{equation}
    \bar{P}_{aa} - \frac{1}{N} = \sum_n \left( p_n^{a}  - \frac{1}{N}\right)^2 \ge 0 \, \Rightarrow \, \bar{P}_{aa} \ge \frac{1}{N}
\end{equation}
due to the normalization $\sum_n p_n^{a} = 1$. Therefore, the quantum dynamics cannot lead to the statistical, ergodic behavior of $\bar{P}_{aa} = 1/N$ unless we start with the artificial state given in Eq.~\eqref{Eq:quantum_ergodic_state}. We argue that the initial configuration $\vert a \rangle $ itself constitutes the enigmatic state responsible for breaking the expectation of full ergodicity. This feature is the fundamental reason behind the ubiquitous factor $P^{\textrm{UQB}}$ in any quantum system.

To further trace the origin of $P^{\textrm{UQB}}$, we present a brief, heuristic overview of the underlying argument. First of all, based on Eqs.~\ref{Eq:time_invariance_of_quantum_measure} and \ref{Eq:dilution_factor}, we can generally conclude
\begin{equation}
    \bar{P}_{aa} = \sum_n p_n^{a}p_n^{a} \ge \sum_n p_n^{a}p_n^{b} = \bar{P}_{ab},
\end{equation}
where $\vert b\rangle$ is dynamically distinguishable but otherwise {\it a priori}  matched to the initial state $\vert a\rangle$. The equality holds only for the initial state $\vert b\rangle = \vert a \rangle$ or any evolute  $\vert b  \rangle  = \vert \alpha \rangle$.  This result stems from the fact that fluctuations in $p_n^{a}$ and $p_n^{b}$ from one $n$ to the next are independent, the squares of Gaussian independent random variables, whereas the state $\vert a \rangle$ obviously correlates with itself. 

{The ideal chaotic system has no symmetries; i.e., the only operators commuting with the Hamiltonian are the identity and the Hamiltonian itself. This means that the spectrum of the Hamiltonian is nondegenerate. Furthermore, the expansion amplitudes $a_n$ in Eq.~\eqref{Eq:expansion_of_the_initial_state} are complex variables, containing 2 degrees of freedom, their real and imaginary parts. In particular, the statistics of the $a_n$ locally follow a $\chi^2(1)$ distribution of 1 degree of freedom. Consequently, we get an amplification of $\bar{P}_{aa}/\bar{P}_{ab} = 2$.
This occurs because every (by chance) large $\bar{P}_{aa}$ gets squared in the sum in Eq.~\eqref{Eq:dilution_factor}, whereas $\bar{P}_{ab}$ has no such systematic enhancement for a typical $\vert b \rangle$ (see also Fig.~\ref{Fig:Pabspectra}). On the other hand, if   time-reversal symmetry applies, the amplitudes $a_n$ can be instead chosen to be real, reducing the number of degrees of freedom to one, consistent with the $\chi^2(1)$ distributions. This small modification yields the slightly higher amplification of $\bar{P}_{aa}/\bar{P}_{ab} = 3$. With additional symmetries, the enhancement factor can seem to be higher, i.e., $\bar{P}_{aa}/\bar{P}_{ab} > 3$, as elaborated to a further extent below and in Sec.~\ref{Sec:time-reversal_and_spatial_symmetry}. For instance, this effect has been contemplated~\cite{Heller_chem.phys.lett_60_338_1979} in the special case of the C3 symmetry associated with the Henon-Heiles potential, which is instead related to the internal dynamics of molecules.   

In summary, we have the universal QB factor of
\begin{equation}\label{Eq:universal_QB}
    P^{\textrm{UQB}} 
    \begin{cases}
        = 2 & \textrm{without time-reversal symmetry},\\
        \\
        = 3 & \textrm{with time-reversal symmetry},\\
        \\
        \ge 3 & \textrm{with additional symmetries}.
    \end{cases}
\end{equation}
Rather than relying on the heuristic argument given above, the result can be derived more rigorously within RMT. In Appendix~\ref{Appx:UQB_derivation}, we present an extended RMT analysis that entails a complete proof of the universal factors in the absence of additional symmetries producing degeneracies in the spectrum, and we further show that the corresponding enhancement persists for higher-order moments. The presence of additional symmetries requires a more subtle treatment. If the considered state is confined to a single symmetry subspace, the familiar RMT enhancement factor of 2 or 3 emerges. The same factor also appears when the state uniformly spans all symmetry-related subspaces. In the intermediate case, however, the overall enhancement relative to the full Hilbert-space dimension typically exceeds the standard RMT value. By contrast, when individual symmetry subspaces are analyzed separately, the usual RMT enhancement is recovered, in agreement with our proposed notion of \emph{a priori} matching discussed above. A more detailed discussion of the role of additional symmetries is deferred to Ref.~\cite{Universal_quantum_birthmark_paper}. Here, we focus on the simplified setting that establishes the general existence of the QB effect, but the influence of time-reversal and spatial symmetries is addressed in Sec.~\ref{Sec:time-reversal_and_spatial_symmetry}.
 
In anticipation of the discussion in Sec.~\ref{Sec:time-reversal_and_spatial_symmetry}, we already highlight an important aspect of time-reversal symmetry. Time-reversal invariance of a Hamiltonian alone is not sufficient to yield the enhancement factor $P^{\textrm{UQB}} = 3$, since the expansion coefficients of the chosen initial state may still be complex, resulting instead in $P^{\textrm{UQB}} = 2$, as demonstrated in our results below. By contrast, if the initial state itself is also time-reversal invariant, the expansion coefficients are necessarily real valued, and the enhancement factor becomes $P^{\textrm{UQB}} = 3$.    

Overall, the UQB is a permanent feature of the initial condition and the dynamics for all time, and is not a transient effect. We want to emphasize that as implied by Eq.~\eqref{Eq:time_invariance_of_quantum_measure}, the time averaging of the probability can be initiated at any time, and still the earlier initial state is remembered, contrary to naive ergodic expectations. This fact also invalidates the ergodicity hypothesis in quantum systems: Long-time averages do not match with ensemble averages, at least up to the UQB. A simple and visually transparent illustration of this point is provided by our stadium results in Sec.~\ref{models}. Whereas the time-averaged probability density obtained from a single wave-packet realization retains a quantum-birthmark structure indefinitely, an average over a sufficiently large stack of eigenstates yields a uniformly distributed probability density. The former constitutes a central result of the present manuscript, while the latter has been established previously~\cite{Antiscarring_1, lu_phys.rev.a_112_043307_2025}, provided that the ensemble of eigenstates spans an energy window exceeding that associated with the shortest periodic orbit of the system.  

Notably, the effect of $P^{\textrm{UQB}}$ does not dilute in the semiclassical limit of $\hbar \rightarrow 0$, as analyzed in Sec.~\ref{Subsection:semiclassical_limit}. We further discuss this weak ergodicity breaking from the perspective of quantum-ergodicity theorems~\cite{Shnirelman_Uspekhi.Mat.Nauk_29_181_1974, Colindeverdiere_comm.math.phys_102_497_1985, zelditch_duke.math.j_55_919_1987} in Sec.~\ref{Subsection:semiclassical_limit}. Additional discussion of the implications of quantum birthmarks for ergodicity and thermalization, along with experimental relevance, is presented in Secs.~\ref{Discussion_A} and~\ref{Discussion_C}, respectively. We also note that the QB phenomenon imposes a rigid constraint on ergodicity in a Hilbert space, thereby amending the concept recently presented in Refs.~\cite{pilatowsky_phys.rev.lett_131_250401_2023, pilatowsky_phys.rev.x_14_041059_2024}. 

\subsection{Revival-enhanced quantum birthmarks}

In a maximally chaotic dynamics determined by RMT~\cite{kaplan_j.phys.A_40_F1063_2007}, the initial state $\vert a \rangle$ and states $\vert \alpha \rangle$ that evolve from it are enhanced in future visitation by a factor of $P^{\textrm{UQB}}$ over the average or typical state. However, the spectral fluctuations of the amplitudes start already hovering around the value of $2/N$ and $3/N$ in the earlier times, depending on the time-reversal symmetry~\cite{pechukas_chem.phys.lett_86_553_1982}. This happens without early recurrences, which we next take into account by introducing revival-enhanced QB. 

We begin with some of the reasons for short-time revivals to occur in systems which can nonetheless be classically ergodic. Assuming the launch of a localized nonstationary wave packet $\vert a\rangle$, its evolution measured by the fidelity $P_{aa}(t)$ may have revivals after the initial nearly complete decay, long before the Heisenberg time $t_H$ when adjacent eigenvalues have dephased. Here are two revival scenarios possible in chaotic systems:
\begin{itemize}
    \item Return of wave-packet amplitude along unstable periodic or homoclinic orbits, albeit  decreasing with repetitions in both situations.
    
    \item A wall bounce or potential feature causing approximate returns; these may lead to strong revivals.

\end{itemize}
We here focus exclusively on (partial) revivals, such as exemplified by the list above, occurring prior to the Heisenberg time $t_H$, which contribute to the revival component of a QB. These should not be confused with possible full quantum revivals, in which the initial state is reconstituted only after the Heisenberg time (see Sec.~\ref{Discussion_A} for further discussion).

A revival happens only when both position and momentum realign, but the fidelity $P_{aa}$ cannot reach unity, because that would lead to a periodic evolution. Nevertheless, it could be very close to the full recovery, making repeat revivals inevitable, in the form of slow exponential decay. In fact, the key to the original discovery and proof of quantum scarring~\cite{Heller_phys.rev.lett_53_1515_1984} was to note that early recurrences cause features to appear in the spectrum of $p_n^a$ vs. $E_n$ with separations greater than the level spacing. As mandated by the normalization $\sum_n p_n^a = 1,$ such oscillations necessitate an increase in $\bar{P}_{aa}$ coming from Eq.~\eqref{Eq:quantum_phase_space_exploration}, which is greater the stronger the recurrences are (see, for instance, Fig.~\ref{Fig:Pab_modulation}).

\begin{figure}[h]
    \centering
    \includegraphics[width=\linewidth]{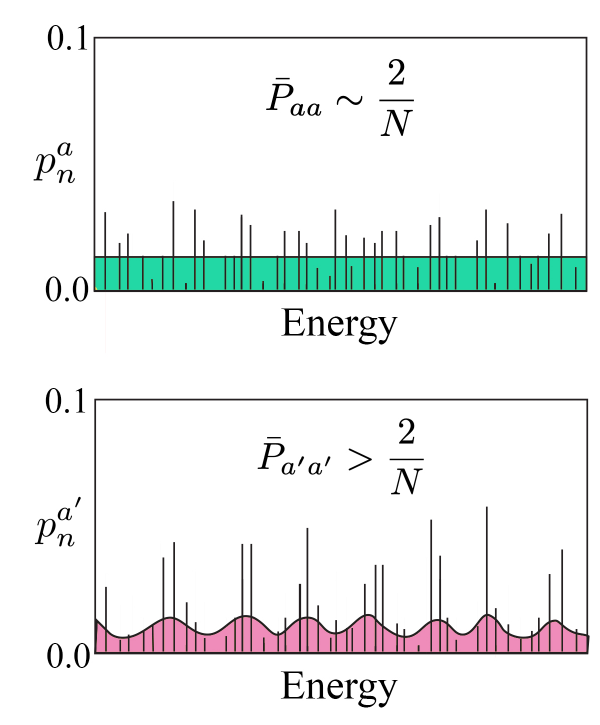}
    \caption{Spectra and long-term occupation of two different initial states $\vert a\rangle$ and $\vert a' \rangle$: The same Hamiltonian $\mathcal{H}$ applies to both states, and therefore they share the same set of eigenvalues. Nonetheless, the individual intensities differ: $\vert a\rangle$ had no early revivals possessing the RMT average (green envelope), but $\vert a' \rangle$ instead underwent one or more revivals, causing the low resolution structure (pink envelope). The spectral lumps for the state $\vert a' \rangle$ mean the intensities must necessarily be distributed more widely, yielding $\bar{P}_{a'a'} > \bar{P}_{aa}$. }\label{Fig:Pab_modulation}
\end{figure}

As stated before, the early-time dynamics, including revivals, has a profound amplification effect on QB. The absence of revival for an initial condition means that there is no simple evidence of dynamics specific to the system. By early, we mean well before the Heisenberg time $t_H$, the time after which entire phase space could have been explored, if kept the maximum exploration rate $\mathcal{R}_{\textrm{max}}$. Such revivals, which may be quite classical in origin, have nonclassical implications for quantum mechanics, caused by constructive and destructive interference. 

Early recurrences in the fidelity $P_{a\alpha}(t)$ cause the spectrum to develop envelope structure that constrains the spectral fluctuations of the components $p_n^{a}$, as displayed in Fig.~\ref{Fig:Pab_modulation}. These constraints cause a failure of ergodicity  with respect to the classical-like and RMT expectations (cf. Eqs.~\ref{Eq:quantum_ergodic_state} and \ref{Eq:universal_QB}, respectively): The ``early-time'' envelope modified by recurrences reveals system-specific dynamics and reduction of phase-space exploration rates. 
These deficiencies are real, in the sense that not as much phase space will be explored as a consequence of the early recurrences. There is always the option of  incorporating these constraints and thus recalibrating the standard for ergodicity. Ergodicity means nothing without stating what the prior constraints are.

In particular, we can allocate the correction of the early dynamics into the revival-enhancement component of a QB. This factor can be estimated in the following way~\cite{Smith_phys.rev.e_80_035205_2009, Smith_phys.rev.e_82_016214_2010} 
\begin{equation}\label{Eq:revival_QB}
    P^{\textrm{RQB}} \simeq  \frac{\int_0^{T^*} P_{a\alpha}(t) dt}{\int_0^{T^*} P_{a\beta}(t)dt} \times \frac{\int_0^{T^*} P_{ab}^{\textrm{RMT}}(t) dt}{\int_0^{T^*} P_{aa}^{\textrm{RMT}}(t)dt},
\end{equation}
where $T^*$ is an appropriately chosen cutoff time and the dynamical information is encoded in the fidelities $P_{a \alpha}$ and $P_{a \beta}$ defined in Eqs.~\ref{Eq:fidelity} and \ref{Eq:generalized fidelity}, respectively. In the RQB, this behavior is then compared against the RMT baseline of
\begin{equation}
\begin{split}
    P_{ab}^{\textrm{RMT}}(t) &= \sum_{n \neq m } \exp\left( i\frac{\varepsilon_n -\varepsilon_m}{\hbar}t \right)\Big[ \vert \langle a \vert \zeta_n \rangle \vert^2 \vert \langle b \vert \zeta_m \rangle \vert^2\\ &+ \langle a \vert \zeta_n \rangle \langle \zeta_n \vert b \rangle \langle \zeta_m \vert a \rangle \langle b \vert \zeta_m \rangle \Big]
\end{split}    
\end{equation}
where energies  $\varepsilon_n$ and basis states $\vert \zeta_n \rangle$ ($n = 1, \dots, N$) are drawn from the appropriate RMT ensemble, such as GOE or GUE, depending on the symmetry class of the system. However, since RMT does not sharply distinguish between different probe states $\vert a \rangle$ and $\vert b \rangle$, the second ratio in the estimator of Eq.~\eqref{Eq:revival_QB} is typically of order unity.

While the approximation above provides only an estimate of the RQB, it becomes exact in three notable limits: (i) when the system exhibits fully universal (RMT-like) dynamics down to the initial time $T^* = 0$, (ii) in the semiclassical limit $\hbar \rightarrow 0$, and (iii) when the entire dynamical evolution is given, i.e., $T^* \rightarrow \infty$. Remarkably, even far from these exact regimes, the estimate remains robust, for instance,in finite-sized systems with weak chaos or when only early-time dynamics on the scale of a few Lyapunov times are available (see, e.g., Refs.~\cite{Smith_phys.rev.e_80_035205_2009, Smith_phys.rev.e_82_016214_2010}).

For the estimation in Eq.~\eqref{Eq:revival_QB}, the required early-time fidelity input can, in some cases, be obtained analytically, for example, regarding the Heller-type~\cite{Heller_phys.rev.lett_53_1515_1984, kaplan_ann.phys_264_171_1998, Kaplan_nonlinearity_12_R1_1999} or many-body scarring~\cite{chandran_annu.rev.condens.Mmatter.phys_14_443_2023}. In more general settings, however, the short-time dynamics for a specific system must be evaluated numerically, up to a desired cutoff time $T^*$. This cutoff is typically chosen to be long enough to capture all nonuniversal features of the dynamics (roughly longer than the Thouless time that is required for an initial state to spread over the available space, before which universal dynamics is not possible), yet short compared to the Heisenberg time, beyond which universal behavior dominates (these timescales usually depend on global symmetries: see, e.g., Refs.~\cite{friedman_phys.rev.lett_123_210603_2019, roy_phys.rev.e_102_060202_2020, roy_phys.rev.e_106_024208_2022}). For a given random matrix ensemble, the corresponding factors can be treated entirely analytically, or alternatively computed numerically. Nevertheless, the optimal choice of $T^*$ does depend on the initial state. For instance, for a wave packet tailored to a periodic orbit that gives rise to the Heller-type scarring, the corresponding revival factor is $P^{\textrm{RQB}} \sim \pi/\chi$, where $\chi$ is the dimensionless stability parameter, defined as the product of the Lyapunov exponent and period of the orbit~\cite{kaplan_ann.phys_264_171_1998, Kaplan_phys.rev.e_59_5325_1999, kaplan_phys.rev.e_62_409_2000}. This qualitative fact lends support to the useful rule of thumb suggested by our observations: The cutoff time $T^*$ could be roughly chosen to be at least a few times longer than the period of the shortest relevant POs~\footnote{In our numerical studies, we used significantly larger values of $T^*$ and verified the stability of the resulting estimates; nevertheless, a systematic determination of the optimal cutoff time remains an open problem.}. 

The existence of an RQB corrects the RMT framework for nonintegrable systems by systematically incorporating short-time dynamics. Crucially, the RQB is fully quantum mechanical; it persists even when semiclassical approximations break down, such as at low energies or in small system sizes. Furthermore, in systems with a classical limit, the RQB correctly accounts for the influence of nonuniversal short POs, which shape early-time dynamics. 

The core insight of the original quantum scarring~\cite{Heller_phys.rev.lett_53_1515_1984} is closely related: A nonstationary initial state can exhibit restricted phase-space exploration due to early partial recurrences or revivals. These recurrences are semiclassical manifestations of moderately unstable POs, which guide the decay and reformation of wave packets. The RQB formalism captures this mechanism and generalizes it beyond the semiclassical domain. Moreover, the notion of an RQB naturally takes into account many-body scarring~\cite{serbyn_nat.phys_17_675_2021}, where the classical limit can be ill-defined or absent. When the initial state overlaps with a scar, early-time revivals lead to a salient RQB. In this sense, RQBs generalize and refine the concept of quantum scarring, shifting emphasis from individual eigenstates to long-time deviations from ergodicity.

However, we stress that the concept of RQB is not limited to quantum scarring; rather, it applies broadly to generic quantum-chaotic systems, whether or not a classical limit exists. This factor corrects the supreme limit of RMT ergodicity. Specifically, we have identified a general form of RQB that originates from any kind of early-time dynamics of the system that is further invariably amplified by a UQB factor. The RQB effect is particularly important for a (narrow) wave packet aligned with a specific trajectory that can be chaotic or periodic, with any existing revivals and recurrences magnifying the phenomenon. These RQB-enhanced features become permanently tattooed on the evolution of a nonstationary state, as we illustrate in the following section.

\section{Unveiling quantum birthmarks} \label{models}

We investigate the ergodicity breaking induced by a QB  in its implications for position space, as opposed to other representations. Both the initial condition $\vert a \rangle$ and its early-time evolution $\vert \alpha \rangle$ leave persistent signatures in the spatial distribution of the long-time average probability density of a quantum particle. 

In Sec.~II, we formulated the QBs in terms of matched states with the same mean energy and energy dispersion. However, the compelling  coordinate-space images presented here in Sec.~III highlight a somewhat different perspective: We can understand the coordinate-space correlation between a wave packet representing the initial state $\vert a \rangle$ and the probe state $\vert b \rangle$ in terms of the Markovian approximation:
\begin{equation*}
    \vert \langle a\vert b \rangle \vert^2  \approx \vert \langle a\vert c \rangle \vert^2  \cdot \vert \langle c\vert b \rangle \vert^2,
\end{equation*}
where the state $\vert c \rangle$ corresponds to a narrow, coordinate-space wavefunction.  Thus, our images serve as convenient -- though indirect -- evidence of the type of $P_{ab}$ correlations discussed previously in Sec.~II. Nevertheless, this can be viewed as effectively integrating over all momenta at a fixed position, since the state $\vert c \rangle$ is assumed to be spatially localized (e.g., a minimum Gaussian). If a state $\vert b \rangle$  is brighter than average at that position, it will be detected, regardless of the direction in which the momentum peaks. This behavior is inferred from the shape of the QB.

\subsection{Model system}

We turn now to the Bunimovich stadium, which is known to exhibit full classical chaos, with every orbit unstable except for the zero-measure, vertical bouncing-ball orbits which are marginal~\cite{Bunimovich_funct.anal.Appl_8_254_1974, Bunimovich_commun.math.phys_65_295_1979}. On the quantum side, it serves as the canonical model in studies of quantum chaos~\cite{Stockmann_book, Heller_book}. For example, it played a pivotal role in the development of the Heller-type scarring theory~\cite{Heller_phys.rev.lett_53_1515_1984, kaplan_ann.phys_264_171_1998, Kaplan_nonlinearity_12_R1_1999} as well as in tests of the BGS conjecture on spectral statistics~\cite{Bohigas_phys.rev.lett_52_1_1984}. 

Commonly, Dirichlet boundary conditions are imposed, forcing the wavefunction to vanish at the stadium walls, as the billiard considered in Sec.~\ref{Subsection:semiclassical_limit}. In addition to adopting such ``hard-wall" boundaries, we also study a softened confinement, where the quantum particle can tunnel slightly into the walls that is a more physically realistic scenario, for instance, in nanoscale devices~\cite{Ge_nature_635_841_2024}. This does presumably cause tiny islands of stability to appear classically that can be smaller than a Planckian cell. Furthermore, there is a more conceptual distinction. For hard-wall billiards, eigenvalues and eigenfunctions can be efficiently accessed utilizing techniques such as the boundary integral method~\cite{kress_math.comp.mod_15_229_1991, backer_2003, veble_new.j.phys_9_15_2007}, as employed in our hard-wall analysis later. While a similar eigenstate-based approach could be, in principle, applied to study QBs in soft confinements, it is typically far more cumbersome in practice. By contrast, the wave-packet approach is always at one's disposal and, in fact, becomes basically indispensable in more complex settings, most notably in many-body systems beyond the scope of this work. This contrast highlights the necessity of a broader paradigm shift away from the eigenstate-centered description toward a time-domain perspective that is also more closely aligned with the tenets of the classical ergodicity.

We start by studying the QB phenomenon in the soft-wall stadium. Instead of an infinite potential barrier analyzed later in Sec.~\ref{Subsection:semiclassical_limit}, we model the boundary as a smooth potential step rising from zero to a finite value $V_0 = 70\,\textrm{eV}$ (see, e.g., Ref.~\cite{Tomsovic_phys.rev.e_47_282_1993} for further justification). The potential is given by
\begin{equation}\label{soft_billiard_potential}
V(\mathbf{r}) = \frac{V_0}{1 + \lambda\exp[\mu(1 - f(\mathbf{r})^2)]},
\end{equation}
where the parameters $\lambda$ and $\mu$ control the softness of the boundary, and the form function $f(\mathbf{r})$ encodes the stadium geometry (the specifics of the simulations are presented in Appendix~\ref{Appx:Wavepacket_simulations}). For the chosen simulation parameters of $\lambda = 25$ and $\mu = 5$, the stadium resembles a bathtub-shaped quantum dot with a flat interior. Classical trajectories with energy below $V_0$ remain largely unaffected by the smooth boundaries. In the limit $V_0/\langle \mathbf{p} \rangle \to \infty$, where $\langle \mathbf{p} \rangle$ denotes the average momentum of the particle, the hard-wall case is even recovered.

Without loss of generality, we initialize the system in a Gaussian wave packet $\vert a \rangle$ centered at $\mathbf{r}_0$, formally
\begin{equation}\label{Initial_Gaussian}
\Psi(\mathbf{r}, 0) = \langle \boldsymbol{r} \vert a\rangle = Z\exp\left[ -\frac{1}{4} \vert (\mathbf{r} - \mathbf{r}_0 ) \cdot \boldsymbol{\sigma} \vert^2 + i \mathbf{k} \cdot \mathbf{r} \right],
\end{equation}
where $Z$ is a normalization factor and $\boldsymbol{\sigma} = (\sigma_x^{-1}, \sigma_y^{-1})$ sets the initial width. The momentum $\mathbf{k} = k(\cos \theta, \sin \theta)$ is characterized by its angle $\theta$ and magnitude $k$ roughly matching with the average energy of the wave packet. This progenitor state is propagated by utilizing the third-order split-operator method (see, e.g., Refs.~\cite{Heller_book, aydin_proc.natl.acad.sci_121_e2404853121_2024, Graf_entropy_26_492_2024, zimmermann_entropy_26_552_2024}). We then compute the autocorrelation 
\begin{equation}
    P_{a\alpha}(t) = \left \vert \int \Psi^*(\mathbf{r}, 0) \Psi(\mathbf{r}, t)\, d\mathbf{r} \right \vert^2,
\end{equation}
and subsequently estimate the averaged occupation probability
\begin{equation}\label{eq:dilation_factor_numerical}
\bar{P}_{aa} \approx \lim_{T\to {\rm large}}\frac{1}{T} \int_{0}^{T} P_{a \alpha}(t) \, dt.
\end{equation}
We already note here that the expected lower bound for the enhancement factor $\bar{P}_{aa}/\bar{P}_{ab}$ 
is 2, as shown below, rather than the factor of 3 that might be naively anticipated based on the time-reversal symmetry of the stadium system. This difference stems from the fact that the initial wave packet defined in Eq.~\eqref{Initial_Gaussian} is complex, and therefore not time-reversal symmetric. On the other hand, when the initial state is instead chosen to be real, the factor of 3 associated with the time-reversal symmetry is recovered, as discussed later in Sec.~\ref{Sec:time-reversal_and_spatial_symmetry}.

Complementary to the $\bar{P}_{aa}/\bar{P}_{ab}$ measure, we define the time-averaged probability density
\begin{equation}\label{Eq:average_density}
\bar{Q}(\mathbf{r}) = A \lim_{T\to {\rm large}}\frac{1}{T} \int_{T_0}^{T_0 + T} \vert \Psi(\mathbf{r}, t) \vert^2 \, dt.
\end{equation}
It is scaled by the area $A$ of the billiard, such that maximal
ergodicity corresponds to $\bar{Q}(\mathbf{r}) = 1$.
We hence call $\bar{Q}(\mathbf{r})$ the scaled time-averaged probability density.
Note that in the case of the soft-wall stadium the area $A$ 
is the classically enclosed area at the mean energy of the wave packet.

In our simulations, we confirmed that the $\bar{Q}(\mathbf{r})$ distribution of a given test state does not depend on the starting time $T_0$, provided that a sufficiently large propagation window $T$ is included. The evolved state $\vert \alpha \rangle$, however scrambled looking, carries memory of what it looked like when it was the state $\vert a \rangle$ in Eq.~\eqref{Initial_Gaussian}, consistent with the time-invariance statement of Eq.~\eqref{Eq:time_invariance_of_quantum_measure}. We checked this robustness against different starting times $T_0 > 0$, chosen arbitrarily away from the reference time $T_0 = 0$. This even included times beyond the Heisenberg time $t_H$, which was estimated to be $t_H \approx 8000$ in the units of the numerical time step $\Delta t$. Furthermore, we verified the convergence of the results by varying $T$ for a given $T_0$. Hereafter, the starting time and simulation time window are set to be $T_0 = 0$ and $T = 7.5\,t_H$, unless stated otherwise.

All the simulation details are available in Appendix~\ref{Appx:Wavepacket_simulations}. Moreover, the key metrics of the QB-related ergodicity breaking are compiled in Table~\ref{Tab:QB_metrics}, which we discuss next in detail (first the soft-wall stadium, and then the parallel hard-wall case in Sec.~\ref{Subsection:semiclassical_limit}).

\begin{table}[h!]
    \centering
    \begin{tabular}{c|ccccc}
       Metric  & Case (A)  & Case (B) & Case (C) & Case (D) \\
       \hline \hline\\
       \multirow{2}{*}{$\max\{\bar{Q}\}$}  & 4.1 & 4.5 & 2.1 & 1.7 \\
         & (4.2) & (4.0) & (1.6) & (1.5) \\\\
       $\max\{\Gamma_S\}$ & 0.41 & 0.48 & 0.50 & 0.51 \\\\
       \multirow{2}{*}{$\bar{P}_{aa}/\bar{P}_{ab}$} & 6.58 & 3.29 & 2.49 & 1.82 \\
        & (8.25) & (3.53) & (2.66) & (1.98) \\\\
       $\max\{\mathcal{N}_{t}\}$  & 675 & 1352 & 3512 & 4741 \\\\
       \multirow{2}{*}{$\mathcal{N}_{\infty}$}  & 677 & 1355 & 3587 & 4897\\
         & (540) & (1264) & (3350) & (4512)\\\\
       \hline \hline
    \end{tabular}
    \caption{The table summarizes the principal quantitative measures employed to assess the degree of QB formation and the associated ergodicity breaking. Unparenthesized values corresponds to the soft-wall stadium given by Eq.~\eqref{soft_billiard_potential}; parenthesized values are computed for the hard-wall counterpart discussed in Sec.~\ref{Subsection:semiclassical_limit}. Here, the enhancement factor $\bar{P}{aa}/\bar{P}{ab}$ for cases (A) and (B) is evaluated taking into account the spatial symmetry of the initial state; without this symmetry reduction, the corresponding values would be larger by a factor of 2. Further details are provided in Sec.~\ref{Sec:time-reversal_and_spatial_symmetry}.
     }
    \label{Tab:QB_metrics}
\end{table}

\subsection{Coordinate space ergodicity}

To shed light on the QB phenomenon, we examine four representative cases. Three of the initial wave packets are launched from the center of the stadium dot, i.e., $\mathbf{r}_0 = (0,0)$, each with a different angle $\theta$ matched with a scar-generating PO: a vertical bouncing-ball orbit with $\theta = 90^{\circ}$ [case (A)], a horizontal straight orbit with $\theta =0^{\circ}$ [case (B)], and a bow-tie orbit with $\theta = 57^{\circ}$ [Case (C)]. In addition, we include a fourth wave packet [case (D)] launched from the upper-right quadrant of the stadium at $\theta = 123^{\circ}$, representing a more generic circumstance where the initial position and momentum direction align with a chaotic trajectory. The initial position and momentum direction of each wave packet are indicated with black arrows in the upper panel of Fig.~\ref{fig:stadium} that displays the computed distributions $\bar{Q}(\mathbf{r})$ for cases (A)–(D).

\begin{figure}[h!]
    \centering
    \includegraphics[width=\linewidth]{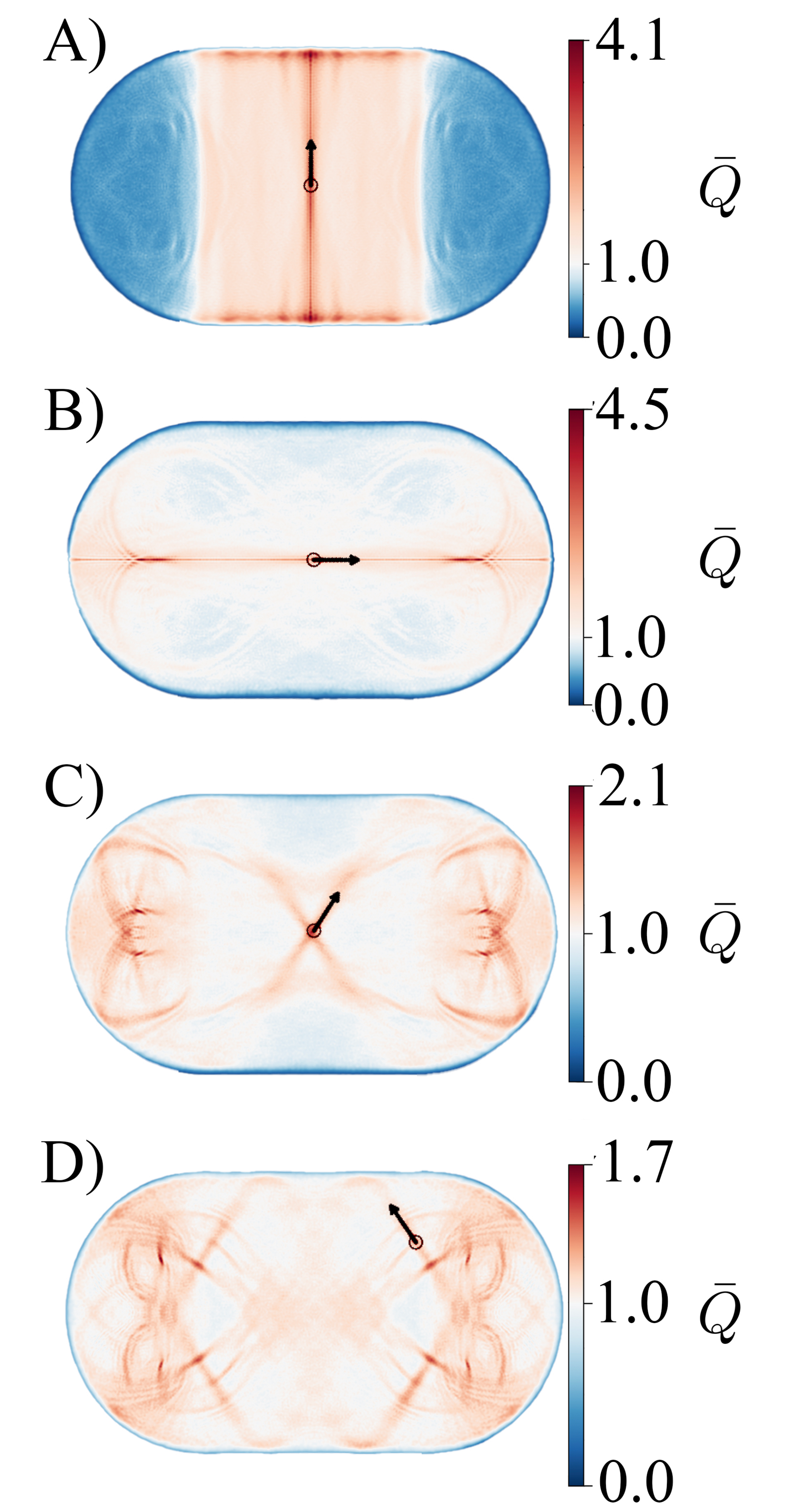}
    \caption{Birthmarks in a soft Bunimovich stadium: Each panel shows the scaled long-time-averaged coordinate-space probability density $\bar{Q}(\mathbf{r})$ of quantum wave packets propagated for cases (A)--(D) specified in the text (corresponding to the panels from the top panel to the bottom, respectively). The strength of early-time recurrences and the associated QB enhancement decreases progressively from top to bottom, as moving from  initial conditions along POs to more generic ones. None of the cases exhibits classical ergodic behavior in the sense of the probability density becoming uniform at infinite time. }
    \label{fig:stadium}
\end{figure}

In all cases seen in Fig.~\ref{fig:stadium}, the distributions of $\bar{Q}(\mathbf{r})$ clearly retain a memory of the early evolution of the wave packets, exhibiting deviations from the ergodicity expected from classical chaos. As mandated by the normalization, any enhancement of probability density in certain regions must necessarily be balanced by a suppression somewhere else. Notably, the resulting distributions respect the two reflection symmetries of the stadium, about both the vertical (minor) and horizontal (major) axes. We also observe significant long-term probability accumulation at focal points, particularly prominent in cases (B) and (C). Moreover, the QB-related enhancements in $\bar{Q}(\mathbf{r})$ reveal structures consistent with caustics formed in the short-time dynamics, especially visible in Cases (C) and (D). This matter is more closely inspected below in the context of Fig.~\ref{fig:sidebyside}. Here, we want to emphasize again that these features of the averaged density $\bar{Q}(\mathbf{r})$ linked to QB formation are invariant with respect to the choice of the starting time $T_0$ employed in the time averaging. This robustness has been both numerically verified and theoretically justified [see Eq.~\eqref{Eq:time_invariance_of_quantum_measure}]. 

In Fig.~\ref{fig:stadium}, regions of increased density (shown in red) appear around the initial state and along its early-time evolution paths, while these enhancements are counterbalanced by suppressed regions (in blue) elsewhere. The magnitude of these enhancements generally correlates with the degree of QB exhibited; see Table~\ref{Tab:QB_metrics}. However, the relationship is not strictly one to one. For example, case (A) exhibits a stronger overall QB signature but spread over a broader area, whereas case (B) shows more spatially localized enhancement in the deviation from the mean value $\langle \bar{Q} \rangle = 1$ (white). This difference can be tracked down to the presence of two focal points in case (B), where the wave packet momentarily accumulates significant density, which is then imprinted into the time-averaged distribution $\bar{Q}(\mathbf{r})$. Such behavior is absent in case (A), resulting in a more diffuse enhancement pattern. Moreover, although cases (C) and (D) exhibit comparable overall levels of QB enhancement, their spatial structures differ: Case (C) associated with a PO displays enhancements that are sharply concentrated along the PO path, while case (D), though linked to a chaotic trajectory, features a more broadly distributed pattern. For further analysis, cross sections of each $\bar{Q}(r)$ plot can be found in Appendix~\ref{Appx:Wavepacket_simulations}.

We emphasize that short-time dynamics play a decisive role in the emergence of the long-term patterns, as expected from the QB argumentation. As shown in Fig.~\ref{fig:sidebyside}, early scattering events leave distinct imprints on the long-time density $\bar{Q}(\mathbf{r})$, marking the presence of a QB. The upper panel displays an enlargement of the lower right quadrant of $\bar{Q}(\mathbf{r})$ of a test Gaussian in the long-time limit, while the lower panel shows the real part of the wave packet shortly after the second wall collision, at $t\approx 0.03\,t_H$, on the same spatial scale. Counterintuitively, nodal patterns formed during this early stage persist to affect the density $\bar{Q}(\mathbf{r})$. Naturally, these features depend sensitively on the initial wave packet, highlighting the ergodicity-breaking nature of a QB. 

\begin{figure}[h]
    \centering
    \includegraphics[width=\linewidth]{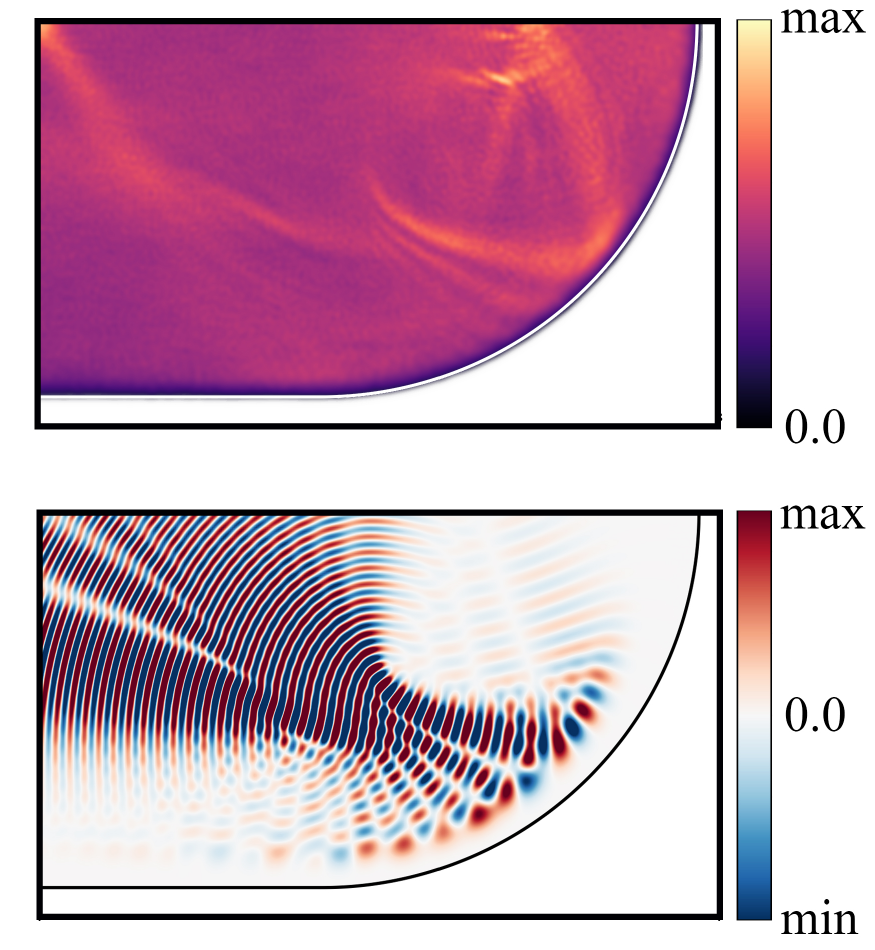}
    \caption{Imprint of the early quantum evolution on the formation of the infinite-time birthmark: The top panel shows the time-averaged probability density in coordinate space $\bar{Q}$ in a magnified region of one quadrant of the stadium dot for case (C). High-intensity areas correspond to regions enhanced by the QB effect. The bottom panel shows the real part of the studied wave packet $\textrm{Re}[\Psi(\mathbf{r},t)]$ shortly after the second bounce off the stadium wall. The early-time wave pattern including the nodal structure lines up well with the localized features observed in the infinite-time average.
   }
    \label{fig:sidebyside}
\end{figure}

We further assess ergodicity in coordinate space by introducing a spatial participation ratio defined as
\begin{equation}
\label{Gamma}
\Gamma_S(t) = A \int \vert \Psi(\mathbf{r}, t)\vert^4  d\mathbf{r},
\end{equation}
where $A$ again refers to the area of the billiard or, more precisely in our case, the area enclosed by the classical turning points at the mean energy of the given wave packet. While the time-averaged dilation $\bar{P}_{aa}$ quantifies participation in the Hilbert space [cf. Eq.~\eqref{Eq:dilation_and_PR}], its spatial analog $\Gamma_S$ captures how uniformly the wave packet spreads across the classically allowed region in coordinate space~\footnote{It is possible to generalize Eq.~\eqref{Gamma}, so as not to be biased toward coordinate space; this can be achieved as follows
\begin{equation}
    \Gamma_\gamma(t) = \frac{1}{\int d \boldsymbol{\gamma}}\int  \ \vert \langle \gamma \vert \Psi(t) \rangle\vert ^4 \, d \boldsymbol{\gamma},
\end{equation}
where $\vert \gamma \rangle $ are members of an overcomplete set $\{\boldsymbol{\gamma}\}$ of coherent states in phase space.
}. 

\begin{figure}[h!]
    \centering
    \includegraphics[width=\linewidth]{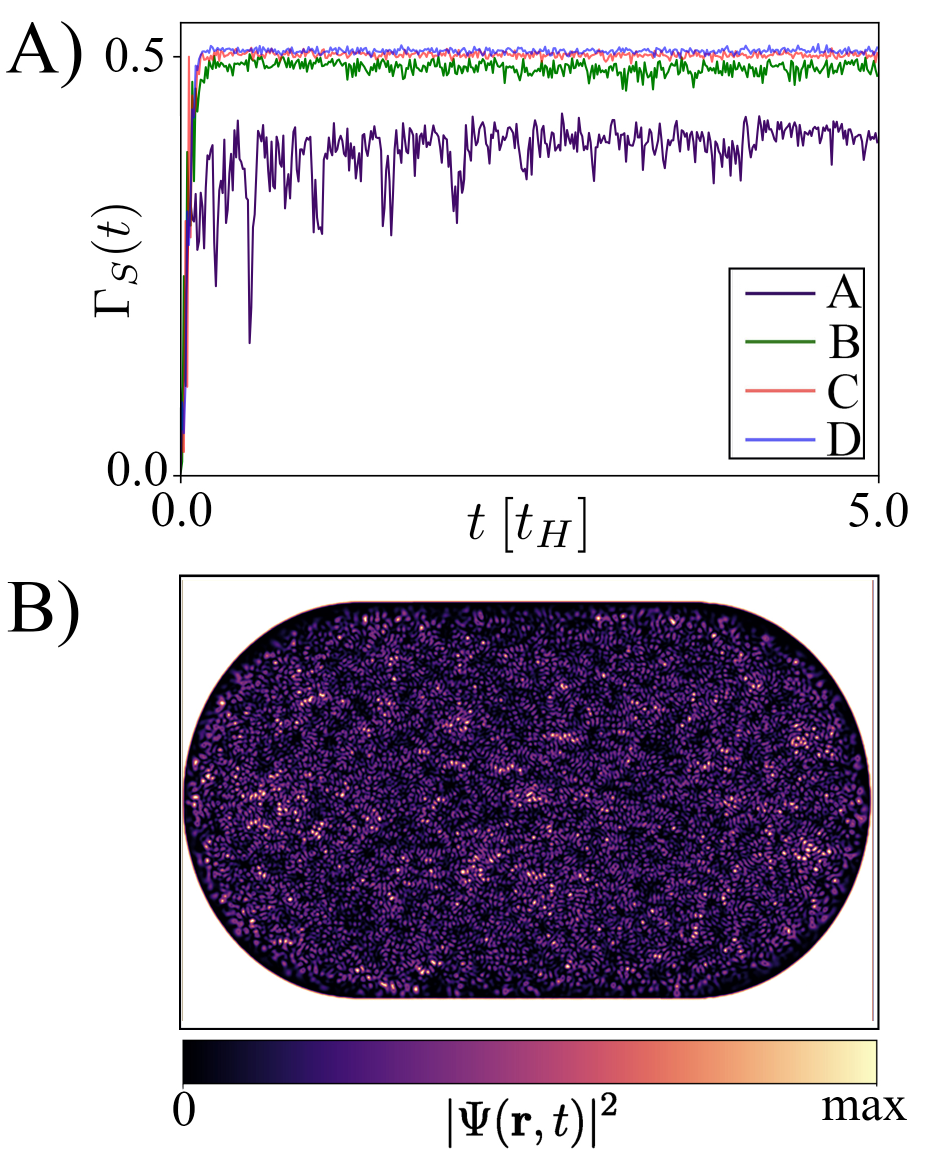}
    \caption{Coordinate space ergodicity: The plot in panel (a) illustrates the instantaneous spatial participation ratio $\Gamma_S(t)$ for the four initial conditions presented in Fig.~\ref{fig:stadium}, serving as a quantitative measure of localization in coordinate space. For all initial conditions of cases (A)-(D), the participation ratio $\Gamma_S(t)$ saturates already after $t \sim 0.1\,t_H$, approaching the value of $1/2$ associated with the Berry-like uniformity of random superposition of plane waves. This action is illustrated in panel (b) showing the probability density of case (D) at time $t= 5\, t_H$. }
    \label{Fig:PRspace}
\end{figure}

As shown in Fig.~\ref{Fig:PRspace}, the defined measure $\Gamma_S$ on spatial uniformity quickly saturates to the value of $\Gamma_S \sim 0.5$ (see Table~\ref{Tab:QB_metrics}) well before the Heisenberg time in all cases studied. At first glance, this seems to contradict with the classical expectation, where ergodicity would suggest $\Gamma_S \sim 1$. However, according to the Berry conjecture~\cite{Berry_j.phys.a_10_2083_1977}, a quantum-chaotic wavefunction approximately resembles a random superposition of plane waves. Such a state naturally yields a quantum expectation of $\Gamma_S \sim 1/2$ due to the local nodal structure, which is in agreement with our observations. Indeed, the wave-packet distributions appear to approach the Berry-like uniformity after only a few reflections, as illustrated in the lower panel of Fig.~\ref{Fig:PRspace}, and further exemplified by the time-evolution animations provided in Ref.~\cite{SM4}. Yet, beneath this apparent randomness lies a Cheshire-cat-like reminiscent of the past, spotlighted through the time-averaged probability density shown in Fig.~\ref{fig:stadium}.

In the same vein, suppose we launch a well-localized wave packet $\Psi(\mathbf{r}, 0)$, with a very small $\Gamma_S$, and evolve it backward in time from a reference time $T_0$ by $-T$, such that the resulting state $\Psi(\mathbf{r}, T_0-T)$ becomes highly dispersed, with $\Gamma_S \sim 0.5$. Suppose we then treat $\Psi(\mathbf{r}, T_0-T)$ as a new initial state. Being delocalized, it carries no clear QB of that moment, except for the UQB. However, assuming that, after evolving forward by a time $T'$, the wave packet may spontaneously collect into a localized state with small $\Gamma_S$. The infinite-time average $\bar{Q}(\mathbf{r})$ then necessarily reveals this event took place at time $T'$.

This thought exercise illustrates a key insight: Spatial QBs correspond to any point in time where the wave packet becomes notably compact, i.e., $\Gamma_S$ becomes small. Crucially, if the wave packet $\Psi(\mathbf{r},t)$ becomes localized again, at any time and place distinct from the original launch at $T_0$, it will leave another detectable mark~\footnote{This argumentation can generalized to the phase-space distribution in terms of the generalized $\Gamma_\gamma$: In the quantum evolution of a classically chaotic Hamiltonian system, the phase-space wave-packet never again, even in the infinite future, spontaneously collects into another compact form with a small value of $\Gamma_\gamma$, for if it did due to symmetry or focus points, that form would also leave its own pattern.}. As seen in Fig.~\ref{fig:stadium}, the observed marks and streaks reflect both the initial state and subsequent dynamics, including transient focusing events caused by symmetries or focal points. The wave packet never spontaneously reassembles into another compact form at any later time; if it did, that event would imprint its own signature in the presented long-time-averaged spatial distribution $\bar{Q}(\mathbf{r})$. However, because of the finite dimensionality of Hilbert space, the original wave packet will eventually reconstruct itself at an extremely long timescale when the phases of all eigenstates return to zero modulo $2\pi$ (see, e.g., Refs.~\cite{styer_am.j.phys_69_56_2001, rasanen_eur_phys.j.b_86_17_2013}).

\subsection{Enhancement factor}

Bridging back to the original discussion of the QB in Sec. II, Fig.~\ref{Fig:autocorr} presents the autocorrelation functions $P_{a\alpha}(t)$ as a function of time for cases (A)–(D). It unmasks a sequence of partial revivals experienced by the studied wave packets prior to the Heisenberg time $t_H$. After the initial decay, governed by the spectral envelope of the wave packet (omitted from Fig.~\ref{Fig:autocorr} for clarity), the strength of the early-time recurrences in the autocorrelations reflects the grade of the associated QB patterns shown in Fig.~\ref{fig:stadium}, which diminishes progressively from case (A) to case (D).

\begin{figure}[h]
    \centering
    \includegraphics[width=\linewidth]{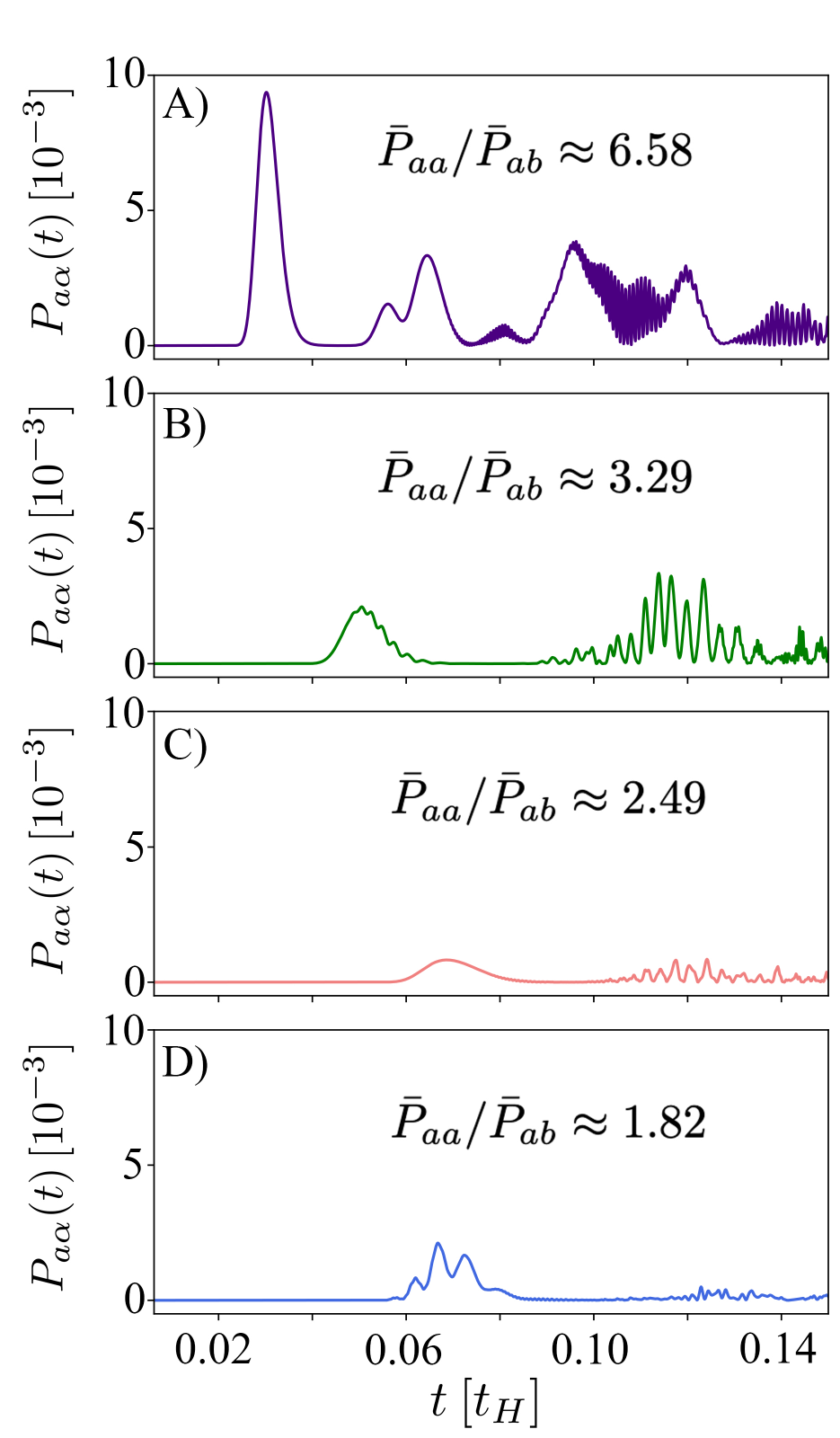}
    \caption{Fidelity and enhancement factor: The figure presents the fidelity for cases (A)--(D) in the short interval near the initial decay, which is omitted for clarity. All cases exhibit revivals taking place before the Heisenberg time $t_H$, which consequently results in a more prominent QB effect, as reflected in the shown enhancement factors $\bar{P}_{aa}/\bar{P}_{ab}$.
    }
    \label{Fig:autocorr}
\end{figure}

The observed revivals are intimately tied to the POs that the wave packets are initially aligned with. For instance, in case (A), the wave packet is synchronized with a marginally unstable bouncing-ball orbit, resulting in strong, persistent revivals as it oscillates between the vertical walls of the confinement before eventually dispersing into the rest of the stadium. In contrast, the wave packet in case (C) broadly falls upon chaotic trajectories; it also partially aligns with a PO contributing to the bow-tie (and diamond) scar pattern visibly in Fig.~\ref{fig:stadium}. This scenery still yields discernible, alas weaker, revivals in its autocorrelation than in case (A). Altogether, these early-time recurrences contribute to the cumulative memory effect characteristic of the QB, in accordance with Eq.~\eqref{Eq:revival_QB}.

In conjunction with the autocorrelation $P_{a\alpha}(t)$, Fig.~\ref{Fig:autocorr} also displays the ratio $\bar{P}_{aa}/\bar{P}_{ab}$, which serves as a measure of the nonergodicity inherent to the QB. These values are also presented in Table~\ref{Tab:QB_metrics}. The numerator $\bar{P}_{aa}$ is obtained by time averaging the autocorrelation function as defined in Eq.~\eqref{eq:dilation_factor_numerical}. For the baseline $\bar{P}_{ab}$, one could average over a large ensemble of dynamically distinct states $\vert b \rangle$, or alternatively, adopt a representative ergodic state associated with a chaotic trajectory. Instead of these two options, we estimate this reference value based on the energy envelope of
the considered wave packets, thus incorporating our notion of an \emph{a priori} constraint, i.e., the probability $p_n^a$ of being in the eigenstate $|E_n\rangle$  depends on energy, as visualized in Fig.~\ref{Fig:Pabspectra}.

More specifically, we model the influence of this wave-packet constraint by decomposing the probability $p_n^a$ into two independent factors
\begin{align}\label{Eq:envelope_decomposition}
    p_n^a = z_n^a \ f(E_n),
\end{align}
where $z_n^a$ is a random variable drawn from the $\chi$ distribution with mean $\langle z_n^a \rangle = 1$, having 1 or 2 degrees of freedom depending on the time-reversal symmetry of the system and the initial state. The function $f(E)$ represents the energy envelope of the wave packet and is normalized according to \mbox{$\int_{-\infty}^\infty dE \, g(E) \, f(E) = 1$}, where $g(E)$ denotes the density of states, ensuring the normalization condition $\sum_n p_n^a = 1$. Next, we want to compare wave packets $\vert b \rangle$ with the same energy envelope $f(E)$. By assuming that $z_n^a$ is independent of $z_n^b$, and combining Eq.~\eqref{Eq:quantum_phase_space_exploration} with Eq.~\eqref{Eq:envelope_decomposition}, we find
\begin{align}\label{Eq:envelope_P_ab}
    \bar{P}_{ab} =  \int_{-\infty}^\infty  g(E) \, f^2(E)\, dE,
\end{align}
and similarly we see that
\begin{align}\label{Eq:envelope_P_aa}
    \bar{P}_{aa} = \langle (z_n^a)^2\rangle \bar{P}_{ab}.
\end{align}
Notably, while the reference value $\bar{P}_{ab}$ itself depends on the chosen energy envelope, i.e., the prior constraint, the enhancement factor $\bar{P}_{aa}/\bar{P}_{ab}$ is independent of it. For example, in the simplest case of a constant energy envelope over a finite-energy window containing $N$ eigenstates, one has $f(E)=1/N$, which yields the expected result of $\bar{P}_{ab}=1/N$.

On the other hand, for a Gaussian energy envelope with width $\sigma_E$, we obtain
\begin{align}\label{Eq:energy_envolope}
    \bar{P}_{ab} = \frac{1}{2 \sqrt{\pi} \sigma_E g}
\end{align}
assuming a constant density of states $g$. This consequently defines an effective Hilbert-space dimension $N = 2 \sqrt{\pi} \sigma_E g$ accessible to wave packets with such an energy profile. Furthermore, for a two-dimensional billiard of area $A$, the density of states may be approximated by the leading Weyl term $g = \frac{A}{4\pi}\frac{2m}{\hbar^2}$,
and the energy width of the Gaussian wave packets considered here is given by $\sigma_E = \frac{\hbar^2 k}{2m \sigma_x}$ with mean momentum $k$ and width $\sigma_x=\sigma_y$ in position space.
Substituting these expressions yields
\begin{align}\label{Eq:energy_envolope_P_ab}
    \bar{P}_{ab} = 2 \sqrt{\pi} \frac{\sigma_x}{A k} \approx \frac{1}{8916},
\end{align}
or equivalently, our effective Hilbert-space size is $N \approx 8916$, with the specific values for $A$, $k$ and $\sigma_x$ given in Appendix~\ref{Appx:Wavepacket_simulations}. We emphasize that the area $A$ is taken as the classical area evaluated at the mean energy $\langle E\rangle$ of the initial wave packet and is assumed constant, as in the hard-wall stadium. While this approximation introduces a small quantitative error, it does not affect the qualitative conclusions regarding the QB effect. This is further corroborated by our hard-wall stadium simulation, which employ the same area $A$ and effective Hilbert-space dimension $N$ and are presented following the soft-wall analysis in Sec.~\ref{Subsection:semiclassical_limit}.

Regarding the wave packet simulations with our soft-wall stadium, the resulting enhancement factor described by $\bar{P}_{aa}/\bar{P}_{ab}$ reveals a striking degree of ergodicity breaking relative to classical expectations (see Table~\ref{Tab:QB_metrics}), even at the UQB level corresponding to the RMT background. This behavior stems from the RQB factor tied to early-time dynamics, particularly the partial revivals of the wave packet $\vert a \rangle$ seen in Fig.~\ref{Fig:autocorr}. The effect is most pronounced in case (A), while in the most ergodic scenario, case (D), the enhancement is close to the expected UQB value of 2. Note that the slightly smaller value of 1.82 is possible for a single wave packet, while the UQB value of 2 is an average. In addition, the values in Table~\ref{Tab:QB_metrics} associated with cases (A) and (B) are symmetry-reduced, as explained in Sec.~\ref{Sec:time-reversal_and_spatial_symmetry}.   

\subsection{Phase-space ergodicity}

In addition, we also examine the emergence of the QB phenomenon from the perspective of quantum phase space, establishing a more concrete link to classical dynamics. In Fig.~\ref{fig:stadium_dyn}, we present the number $\mathcal{N}_t$ of phase space cells visited by the wave packet over time $t$, as defined in Eq.~\eqref{Eq:accessed_phase_space_cells} (see also Table~\ref{Tab:QB_metrics}). This measure shows a saturation for all selected initial conditions, indicating a bounded exploration of the available phase space. The limiting values stated in Table~\ref{Tab:QB_metrics} are approximately given by the Hilbert-space-based estimate
\begin{equation}\label{eq_phase_space_limiting_value}
\mathcal{N}_{\infty} \sim \frac{\bar{P}_{ab}}{\bar{P}_{aa}} N,
\end{equation}
underscoring that, because of the QB effect, the wave packets fail to explore the full phase space, unlike the classical chaos dynamics.

\begin{figure}[h]
    \centering
    \includegraphics[width=\linewidth]{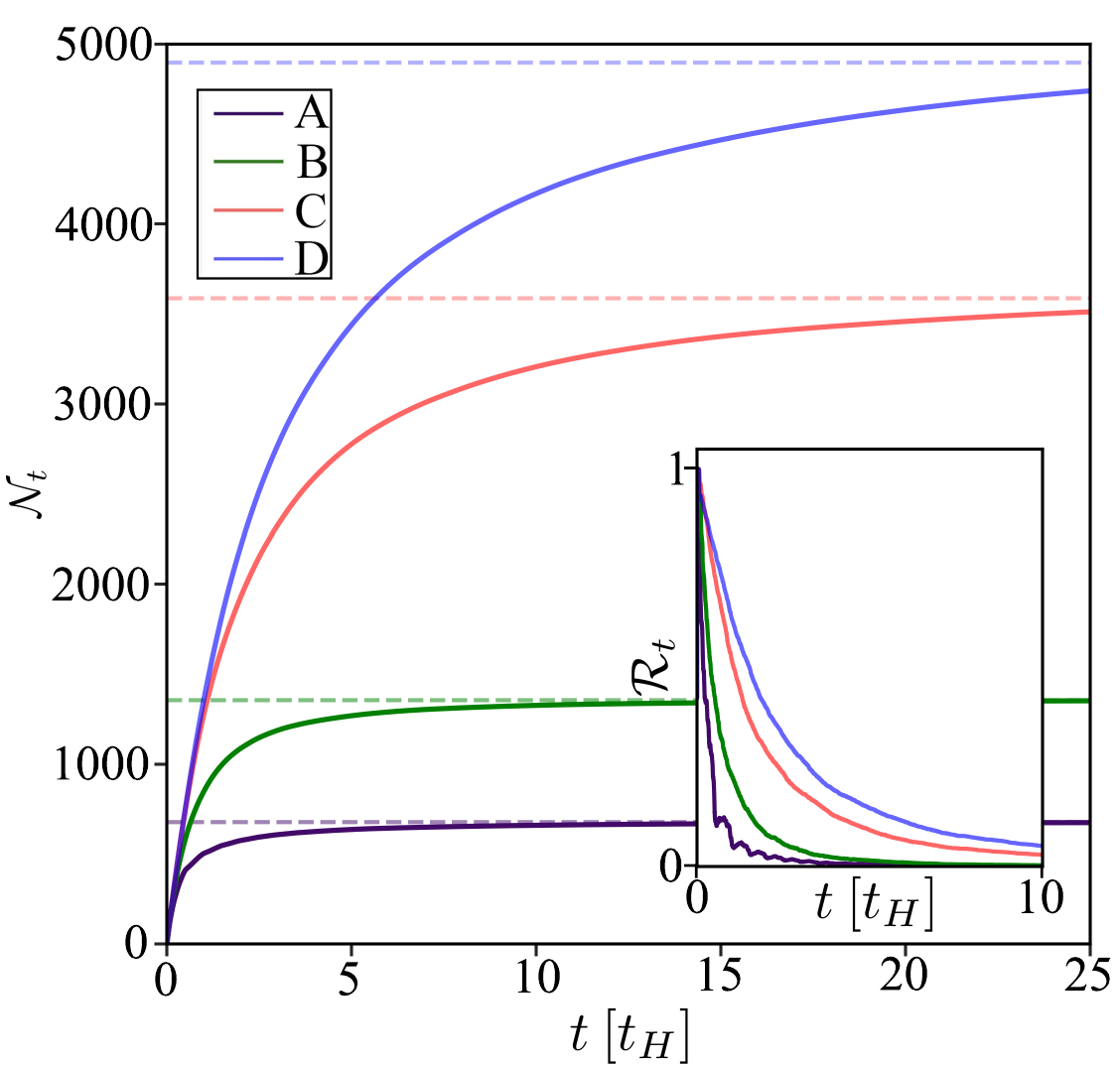}
    \caption{Phase space exploration: The time evolution of the number of explored phase-space cells $\mathcal{N}_t$ is shown for the selected initial conditions (A)--(D). In all cases, $\mathcal{N}_t$ saturates at the timescale beyond the Heisenberg time. However, due to the QB phenomenon, it reveals a bounded exploration of the nominally viable phase space corresponding to $\mathcal{N}_t \sim N \approx 8900$ as $t \rightarrow \infty$. The dashed lines represent the estimated saturation values $\mathcal{N}_\infty$ determined by Eq.~\eqref{eq_phase_space_limiting_value}.}
    \label{fig:stadium_dyn}
\end{figure}

For example, the wave packet in case (A), closely tracking a bouncing-ball orbit (as reflected in its autocorrelation in Fig.~\ref{Fig:autocorr}), saturates at a relatively small $\mathcal{N}_{\infty}$ corresponding to only a narrow fraction of the entire phase space, whereas its classical analog would eventually explore the accessible phase space more uniformly. Strikingly, even in the most ergodic scenario, case (D), the wave packet manages to cover only about one half of the nominal phase space. This highlights a strong correspondence between ergodicity breaking in the abstract Hilbert space and its geometric manifestation in quantum phase space. 

Nonetheless, there is a notable distinction that lies in the timescales involved. While the spectral components of the wave packet are resolved with near-complete accuracy by the Heisenberg time, the number of accessed phase-space cells $\mathcal{N}_t$ continues to grow. This behavior arises from interference effects in the survival probability and the softening in definition of Eq.~\eqref{Eq:accessed_phase_space_cells}. Yet both influences diminish in the long-time limit $t \gg t_H$, where $\mathcal{N}_t$ converges to a value consistent with the ergodicity estimate based on the enhancement factor $\bar{P}_{aa}/\bar{P}_{ab}$ of the averaged occupation probability.

The limited phase space access is also evident in the exploration rate $\mathcal{R}_t$ defined by Eq.~\eqref{eq:phase_space_rate}. As shown in Fig.~\ref{fig:stadium_dyn}, none of the wave packets maintain the maximum possible rate $\mathcal{R}_{\textrm{max}}$ determined in Eq.~\eqref{eq:phase_space_rate_maximum}; instead, they exhibit a steady decline. This decline originates from autocorrelation revivals illustrated in Fig.~\ref{Fig:autocorr}. We find that a faster decay of $\mathcal{R}_t$ from its theoretical maximum correlates with a more restricted overall phase-space exploration. This observation supports the proposed maximum rate principle, thus further linking early-time dynamics to long-time ergodicity limitations.

\subsection{Semiclassical limit} \label{Subsection:semiclassical_limit}

We next turn our attention from the QBs in the soft-wall stadium to its hard-wall counterpart, with the same area $A$ as in the studies above. This line of inquiry serves a dual purpose: first, to validate the soft-wall results, and second and more importantly to examine the persistence of QBs in the semiclassical limit $\hbar \to 0$ in a simplified setting that avoids complications arising from boundary softness. For instance, the classical dynamics of a hard-wall billiard is energy independent and fully ergodic. All eigenfunctions of this system are computed via the boundary integral method~\cite{kress_math.comp.mod_15_229_1991, backer_2003, veble_new.j.phys_9_15_2007} up to a dimensionless wave number $kR = 350$ ($k =\sqrt{2mE}/\hbar $ and $R$ denotes the radius of the semicircular end caps of the stadium), employing the same spatial resolution as in the soft-wall stadium calculations discussed above (see also Appendix~\ref{Appx:Wavepacket_simulations}).

In this setting, rather than explicitly propagating a wave packet in time, we expand the initial nonstationary state $\lvert a \rangle$ in the solved eigenbasis $\{\lvert E_n \rangle\}$. Using this expansion, the infinite-time average of the coordinate-space probability density can be evaluated analytically. For the scaled time-averaged density $\bar{Q}(\mathbf{r})$, we find 
\begin{equation}
\label{Eq:average_density_eigenfunctions}
\bar{Q}(\mathbf{r})
= A \sum_n p_n^{a} \, \big| \langle \mathbf{r} \vert E_n \rangle \big|^2 ,
\end{equation}
where $p_n^{a} = |\langle E_n \vert a \rangle|^2$ denotes the spectral weight of the initial wave packet on the eigenstate $\lvert E_n \rangle$. This quantity is indeed the eigenstate counterpart to the wave-packet definition in Eq.~\eqref{Eq:average_density}: The long-time-averaged density $\bar{Q}(\mathbf{r})$ is thus a weighted superposition of eigenfunction probability densities, with weights entirely determined by the initial condition. Because of the normalization $\langle E_n \vert E_n \rangle = 1$, this choice, as well as Eq.~\eqref{Eq:average_density}, ensures that perfect (classical) ergodicity corresponds to a uniform distribution $\bar{Q}(\mathbf{r}) = 1$. Furthermore, we can similarly compute the enhancement factor $P_{aa}/P_{ab}$ according to Eqs.~\ref{Eq:envelope_P_aa} and \ref{Eq:envelope_P_ab} for the given prior constraint (the same Gaussian energy envelope with the width $\sigma_E$ as above).

Figure~\ref{fig:semiclassical} illustrates the QB for increasing wave number $k$ for the same initial wave packet [case~(D)] previously shown for the soft-wall stadium in Fig.~\ref{fig:stadium}. The central panel corresponds to the wave number employed in the soft-wall calculations of Fig.~\ref{fig:stadium}. The close visual agreement between the corresponding panels demonstrates that the resulting QB patterns are essentially identical in soft- and hard-wall stadiums. This confirms that the observed QB features are insensitive to our choice of confinement softening. Further quantitative agreement is provided by the QB metrics, namely, $\max\{\bar{Q}\}$, $P_{aa}/P_{ab}$, and $\mathcal{N}_{\infty}$, which are computed for all cases (A)–(D) in the hard-wall stadium and presented in Table~\ref{Tab:QB_metrics} alongside the corresponding soft-wall results.

\begin{figure}[h]
    \centering
    \includegraphics[width=\linewidth]{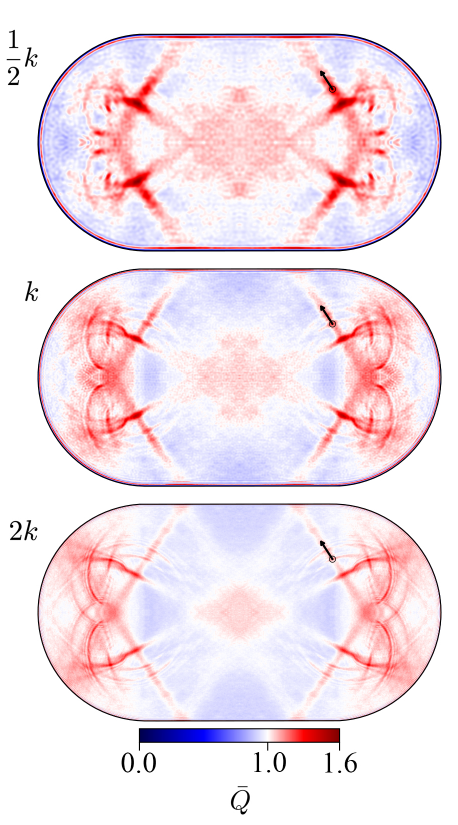}
    \caption{
            The figure presents three scaled time-averaged probability densities $\bar{Q}(\mathbf{r})$ according to Eq.~\eqref{Eq:average_density_eigenfunctions} for a hard-wall Bunimovich stadium with increasing wave number (top to bottom). The stadium geometry is analogous to the soft-wall case studied previously, and the central panel corresponds to the soft-wall stadium value of $k$ of case (D) shown in Fig.~\ref{fig:stadium}. For better comparison, the width of the wave packet is kept fixed (visualized by a circle) and the same color bar is used for all figures.}
    \label{fig:semiclassical}
\end{figure}

Figure~\ref{fig:semiclassical} further unveils that the QB pattern develops progressively finer spatial structure as the wave number of the initial wave packet increases (from top to bottom). This behavior is fully expected: According to Eq.~\eqref{Eq:average_density_eigenfunctions}, the long-time-averaged density $\bar{Q}(\mathbf{r})$ is constructed from eigenfunctions contributing at the energy scale set by the wave packet, and higher energies correspond to shorter de~Broglie wavelengths, leading to finer spatial variations in the eigenfunction probability densities. Consistent with this picture, the numerical data reveal that the characteristic spatial scale of the QB features decreases with increasing wave number. A preliminary analysis based on simulations performed at fixed initial wave packet width $\sigma_x=\sigma_y$ (see Eq.~\eqref{Initial_Gaussian}) suggests that this scale is approximately compatible with a $1/\sqrt{k}$ dependence, which is larger than $1/k$ scaling of the de Broglie wavelength. We have also verified that the same qualitative behavior persists when the width of the initial wave packet is reduced concurrently with increasing $k$ in the semiclassical regime (not shown), indicating that the refinement of the QB structure is controlled primarily by the energy scale rather than by the specific choice of initial wave packet width. 

In addition, we observe that the magnitude of the deviation from ergodicity in coordinate space, quantified by the deviation of $\bar{Q}(\mathbf{r})$ from unity, decreases gradually as the semiclassical limit is approached. This trend is evident in Fig.~\ref{fig:semiclassical} as a progressive reduction in contrast at larger wave numbers, manifested by increasingly lighter red and blue regions from the top to the bottom panel. Furthermore, Table~\ref{Tab:QB_metrics_semiclassical} shows that the quantity $\max\{\bar{Q}\}$ decreases slowly as the semiclassical limit is approached, consistent with the visual observation. In evaluating this maximum, however, we restrict the analysis to a region located a few wavelengths away from the boundary in order to avoid contamination from Friedel-like oscillations near the hard-wall boundary, similar to what is reported in Refs.~\cite{backer_phys.rev.e_57_5425_1998, backer_phys.rev.e_80_066210_2009}. Without this restriction, the measure $\max\{\bar{Q}\}$ would be dominated by the boundary-induced oscillations, yielding spurious values that persist in the semiclassical limit but are unrelated to the QB effect. This perceived fading of the QB in coordinate space is further characterized by the standard deviation $\sigma_{\bar Q}$ of $\bar{Q}(\mathbf{r})$ that gradually decreases in the semiclassical limit, as shown in Table~\ref{Tab:QB_metrics_semiclassical}.

At first glance, this behavior in coordinate space may seem to contradict our prior claim regarding the persistence of the QB effect in the semiclassical limit. However, the reduction of QB signatures in coordinate space is a consequence of projecting onto position alone. As the wave number increases, a fixed spatial resolution effectively averages over an increasing number of Planck cells in phase space. This coarse graining smooths out local fluctuations, resulting in a reduced apparent deviation from ergodicity in $\bar{Q}(\mathbf{r})$, even though the underlying QB enhancement remains intact. In fact, the enhancement factor $P_{aa}/P_{ab}$ remains essentially invariant as a function of the energy, as computed for the representative case shown in Fig.~\ref{fig:semiclassical} and reported in Table~\ref{Tab:QB_metrics_semiclassical}. Consistent with this viewpoint and pointed out earlier, the QB factors of $P^{\textrm{UQB}}$ and $P^{\textrm{RQB}}$ do not explicitly involve Planck’s constant $\hbar$, thus implying the robustness even at the semiclassical limit of $\hbar \rightarrow 0$. In other words, the QB phenomenon and its associated patterns persist at all energies, while their characteristic spatial structures are pushed to progressively smaller length scales in coordinate space as the wave number increases. 
 
\begin{table}
    \centering
    \begin{tabular}{c|ccc}
       Energy & $P_{aa}/P_{ab}$ & $\max\{\bar{Q}\}$ & $\sigma_{\bar Q}$ \\
         \hline \hline\\
        $k/2$ & 1.83 & 1.56 & 0.133\\\\
        $k$ & 1.98 & 1.45 & 0.109\\\\
        $2k$ & 2.04 & 1.35 & 0.085\\\\
         \hline \hline
    \end{tabular}
    \caption{The table summarizes the QB metrics of the enhancement factor $P_{aa}/P_{ab}$, the maximum time-averaged density $\max\{\bar{Q}\}$ (excluding a few wavelengths near the boundary), and the standard deviation $\sigma_{\bar Q}$ of $\bar{Q}$ 
    for case (D) at different energies, characterized by the waven umber. The value marked by $k$ corresponds to that used for the soft-wall stadium simulations (see Fig.~\ref{fig:stadium}).
    Although the $\max\{\bar{Q}\}$ measure and the standard deviation $\sigma_{\bar Q}$ suggest a progressive attenuation of the QB in coordinate space, the associated enhancement factor in Hilbert space remains effectively unaffected as a function of the energy.
    }
    \label{Tab:QB_metrics_semiclassical}
\end{table}

The semiclassical analysis of the QB phenomenon above illuminates a subtle but important aspect. The ergodicity theorems~\cite{Shnirelman_Uspekhi.Mat.Nauk_29_181_1974, Colindeverdiere_comm.math.phys_102_497_1985, zelditch_duke.math.j_55_919_1987} assert that ``most'' high-energy eigenstates become evenly spread out over the available phase space, or according to the Berry conjecture~\cite{Berry_j.phys.a_10_2083_1977} maximally random, subject only to the constraint of energy conservation and other symmetries. Our notion of QB transcends this eigenstate perspective by reinstating dynamics into the consideration on quantum ergodicity, thereby also reconnecting with the classical picture. While eigenfunctions contribute to the QB effect, it cannot be completely inferred from their properties alone, as the phenomenon also involves the structure of the initial state, specifically, the relative phases and correlations among the eigenstates composing it. Even if individual eigenstates are equidistributed in phase space, or uniform in coordinate space up to quantum fluctuations, the fluctuating phases and amplitudes entering the unitary time evolution generically give rise to a nonuniform infinite-time average. In this sense, eigenstates can fully satisfy the quantum-ergodicity theorems, while the QB effect persists, even in the semiclassical limit. We underscore that the QB phenomenon is not rooted in a breakdown of eigenstate equidistribution, but rather in their correlations.

In this respect, the QB effect bears resemblance to quantum scarring. While scarred eigenstates persist even in the limit $\hbar \to 0$, their fraction among all eigenstates vanishes in that limit. By contrast, the QB effect also persists and is, in fact, insensitive to the semiclassical limit at the level of Hilbert space, even though its overall weight in phase space becomes increasingly diluted, much like the coordinate-space projection illustrated in Fig.~\ref{fig:semiclassical}. Nevertheless, the QB effect should remain locally robust, becoming only globally negligible in phase space. A systematic phase-space characterization of the QB effect therefore constitutes a natural and important direction for future research, for example, in the context of Baker's map~\cite{oconnor_ann.phys_207_218_1991}. This is also expected to further elucidate the broader implications of the QB phenomenon, such as discussed in Sec.~\ref{Sec:discussion}.

While we have discussed the QB effect in the semiclassical limit $\hbar \to 0$, it is important to distinguish another relevant, semiclassical-like limit, namely, $N \rightarrow \infty$ at fixed Planck’s constant, which is sometimes referred to as a ``thermodynamic'' limit. However, we eschew this terminology here, as the considerations apply equally to single-particle and many-body systems with a large number of degrees of freedom. 

From a pragmatic viewpoint, while it is possible to design and study quantum systems with relatively small Hilbert-space dimensions nowadays, initial-state engineering provides an additional and functional way to control the effective dimensionality relevant to the dynamics. For example, the effective Hilbert space can be tailored by adjusting the width of a Gaussian wave packet [see Eqs.~\eqref{Eq:energy_envolope} and \eqref{Eq:energy_envolope_P_ab}]. Therefore, although the Hilbert space of the stadium examples above is formally infinite, the prior QB analysis can be confined to effective subspaces of moderate sizes of $N_{\mathrm{eff}} \sim 10^4$ [see Eq.~\eqref{Eq:energy_envolope_P_ab}]. In particular, this approach enables a controlled exploration of QB effects by restricting the system to a suitably, chosen-sized subspace, facilitating a systematic investigation of the limit $N \rightarrow \infty$

Theoretically, in contrast to the semiclassical limit $\hbar \to 0$ where the QB effect remains robust simply due to the $\hbar$ independence [cf. Eqs.~\ref{Eq:quantum_phase_space_exploration_density_matrix} and \ref{Eq:dilution_factor}], the visiting probabilities $\bar{P}_{aa}$ and $\bar{P}_{ab}$ scale nominally as $\sim 1/N$. This behavior seems to suggest the demise of the QB effect as $N \to \infty$. Nevertheless, both factors decay at the same rate, so that their ratio $\bar{P}_{aa}/\bar{P}_{ab} \ge 2$ remains finite. At the minimum, the universal QB enhancement factor $P^{\textrm{UQB}}$ persists unchanged in the limit $N \to \infty$. Moreover, the overall QB factor of $\bar{P}_{aa}/\bar{P}_{ab} \simeq P^{\textrm{UQB}} P^{\textrm{RQB}}$ is often further amplified by the revival factor $P^{\textrm{RQB}}$. 

Generally speaking, the fading of observable QB signatures is typically much slower than the straightforwardly assumed fade of $1/N$, true even in coordinate space as illustrated in Fig.~\ref{fig:semiclassical} and Table~\ref{Tab:QB_metrics_semiclassical}. This fact accentuates that the effect remains experimentally relevant even in large systems, particularly in single-particle realizations, which is showcased in Sec.~\ref{Discussion_C}. At first glance, this endurance of the QB effect in the semiclassical limits $\hbar \to 0$ and $N \to \infty$ may appear to be in tension with the standard understanding of eigenstate thermalization. However, as we discuss in Sec.~\ref{Discussion_A}, no such conflict arises; rather, the QB framework provides additional insight into the underlying circumstances behind ETH~\cite{Srednicki_phys.rev.a_50_888_1994, Deutsch_phys.rev.a_43_2046_1991}.

\subsection{Role of time-reversal and spatial symmetries}\label{Sec:time-reversal_and_spatial_symmetry}

Finally, we briefly discuss the influence of time-reversal and spatial symmetries on the QB phenomenon in our simulations. Our Hamiltonian is time-reversal symmetric, implying that its eigenfunctions can be chosen to be real. This property alone, however, does not guarantee that the expansion coefficients of a generic initial state are also real. This is indeed the case for the Gaussian wave packet considered in Eq.~\eqref{Initial_Gaussian}, whose expansion coefficients are complex, and therefore the expected lower bound for the enhancement factor is the universal value $P^{\textrm{UQB}} = 2$. In this situation, the coefficients may be regarded as drawn from a $\chi$ distribution with 2 degrees of freedom, as discussed above. By contrast, if the initial state itself is also time-reversal symmetric, for instance, by starting instead of Eq.~\eqref{Initial_Gaussian} with the akin real wave packet 
\begin{equation}\label{Initial_Gaussian_real}
\Psi(\mathbf{r}, 0) = \tilde{Z} \exp\left[ -\frac{1}{4} \vert (\mathbf{r} - \mathbf{r}_0 ) \cdot \boldsymbol{\sigma} \vert^2  \right] \cos \left( \mathbf{k} \cdot \mathbf{r} \right),
\end{equation}
the expansion coefficients are necessarily real. In that case, the corresponding $\chi$ distribution has a single degree of freedom, and the universal bound is increased to $P^{\textrm{UQB}} = 3$. We have numerically verified this behavior for case (D) in the hard-wall stadium. On the  other hand, if the Hamiltonian is not time-reversal symmetric, for example, due to the presence of an external magnetic field, the eigenfunctions cannot be chosen to be real, and the expansion coefficients of a generic state are necessarily complex. 

Second, we note that the fourfold symmetry observed in the time-averaged distributions $\bar{Q}(\mathbf{r})$ in Figs.~\ref{fig:stadium} and~\ref{fig:semiclassical} follows directly from the symmetry properties of the individual eigenfunctions, which respect to the two reflection symmetries of the stadium. Beyond the symmetries of the eigenfunctions themselves, the symmetry of the initial state also plays an essential role, analogous to the case of time-reversal symmetry discussed above.

Earlier, we accounted for the constraint imposed by the energy envelope of the Gaussian wave packet when determining the effective Hilbert-space dimension. However, in cases (A) and (B), the initial wave packet is centered on a spatial symmetry line. As a result, the effective Hilbert space is basically reduced by a factor of 2. We have explicitly verified for cases (A) and (B) of the hard-wall stadium that only half of the eigenstates within the effective Hilbert space carry nonzero expansion coefficients in these cases. By contrast, in cases (C) and (D), the initial wave packet explores the full effective Hilbert space. In practice, this implies that the enhancement factor can be evaluated employing the effective Hilbert-space dimension determined by Eq.~\eqref{Eq:energy_envolope_P_ab}, but reduced by a factor of 2 to account for the spatial symmetry in cases (A) and (B). In essence, we analyze the emergence of a QB within a symmetry-closed subspace. Within this symmetry-reduced framework, the universal enhancement factor $P^{\textrm{UQB}}$ can take only the values 2 or 3, depending on the time-reversal symmetry of the Hamiltonian and the initial state.

If the enhancement factor is instead defined with respect to the full effective Hilbert space, the QB effect appears superficially stronger. Analogous to the role of the energy envelope, spatial symmetries may be regarded as an \emph{a priori} constraint on the quantum dynamics of the state. At the theoretical level, their influence can be absorbed into the definition of the universal factor $P^{\textrm{UQB}}$: When all relevant \emph{a priori} constraints (symmetry and envelope) are properly accounted for, the symmetry-reduced description always yields the RMT bounds $P^{\textrm{UQB}}=2$ or $3$. Alternatively, if symmetry reduction is not performed, additional symmetries typically lead to an apparent enhancement exceeding the RMT value, i.e., $P^{\textrm{UQB}}>3$. A more detailed discussion of the role of additional symmetries is presented in Ref.~\cite{Universal_quantum_birthmark_paper}.

\section{Discussion and outlook} \label{Sec:discussion}

The concept of QB stands in contrast to time-independent approaches to quantum ergodicity and chaos, which are primarily based on eigenstates and energy spectra (see, e.g., Refs.~\cite{Gutzwiller_book, Haake_book, Tabor_book, Nakamura_book, Casati_book, Stockmann_book}). Although the apparatus of the time-independent Schr{\"o}dinger equation is always available, relying on eigenstates becomes highly impractical outside of idealized and simple circumstances. In contrast, wave-packet-based approaches avoid this limitation, and time-varying fields and potentials virtually demand it (see, e.g., Refs.~\cite{Heller_book, kim_phys.rev.b_106_054311_2022, aydin_proc.natl.acad.sci_121_e2404853121_2024, aydin_proc.natl.acad.sci_122_e2426518122_2025}). Moreover, our QB framework presented in Sec.~II applies to interacting many-body systems where the eigenfunction approach becomes cumbersome. 

As clearly seen in our stadium examples above, and further illustrated in Fig.~\ref{fig:promo}, the QB signals the demise of ergodicity crucial to classical chaos. Our QBs underscore a shift in quantum thermalization issues from stationary to nonstationary states, thus bringing the quantum treatment more in line with classical mechanics.

\subsection{Meaning for ergodicity and thermalization}\label{Discussion_A}

In our discussion on QBs here we have encountered a neglect of the role of prior knowledge when assessing ergodicity. In classical systems, dynamics may appear fully ergodic once constraints from conserved quantities are accounted for. A similar idea holds in quantum mechanics~\cite{von_Neuman_ergodicity, pechukas_chem.phys.lett_86_553_1982}: All quantum systems are ergodic except those with degenerate subspaces. Just as classical ergodicity is defined relative to conservation laws, quantum ergodicity must be framed with respect to symmetries with the choice of random matrix ensemble, which defines the UQB factor and sets the upper bound of quantum ergodicity. This connection, particularly its implications for wavefunction statistics, has received little attention. The UQB governs properties like transition probabilities, participation ratios, purity, and von Neumann entropy~\cite{kaplan_j.phys.A_40_F1063_2007}, independent of specific spectral features. As such, it applies across ensembles with matching symmetry classes; for example, the UQB derived from the Gaussian orthogonal ensemble also applies to any time-reversal-invariant ensemble, including circular ones.

The pure RMT approach deliberately excludes system-specific details in pursuit of universality. Deviations from its predictions, such as eigenstate scarring~\cite{Heller_phys.rev.lett_53_1515_1984, keski-rahkonen_phys.rev.lett_123_214101_2019}, are commonly attributed to boundaries~\cite{Bies_j.phys.A_36_1605_2003}, finite size~\cite{urbina_eur.phys.j.spec.top_145_255_2007}, diffusive transport~\cite{mirlin_phys.rep_326_259_2000}, or interactions~\cite{kaplan_phys.rev.lett._84_4553_2000}. Similarly, deviations from RMT spectral statistics arise from nonuniversal short-time dynamics~\cite{seligman_phys.rev.lett_53_215_1984, Berry_proc.r.soc.lond.a_400_229_1985}. The concept of an RQB refines the UQB by incorporating these short-time dynamical effects, providing a more accurate account of ergodicity, for instance, in the aforementioned situations.

\begin{figure}[h!]
    \includegraphics[width=1\linewidth]{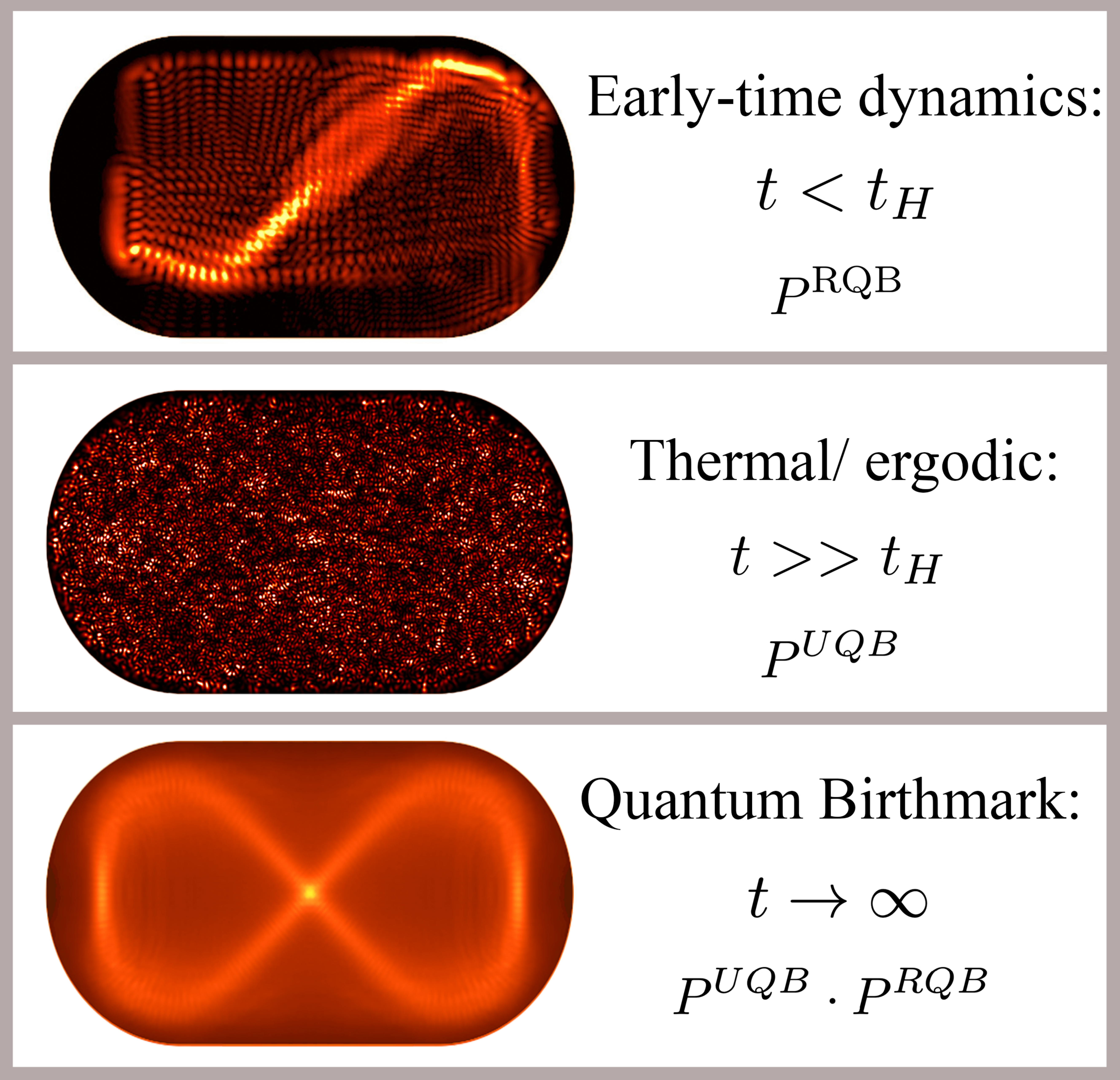}
    \caption{Interplay of universal and revival-enhanced quantum birthmarks: Figure illustrates the core principle behind QB physics. The early dynamics $t < t_H$ of the wave packet (top), such as recurrences, is encoded in the revival factor $P^{\textrm{RQB}}$. At later times ($t \gg t_H$), the system approaches an ergodic or thermal-like distribution (middle), yet still retains a universal imprint of the initial state through the memory factor $P^{\textrm{UQB}}$. These two components together give rise to the QB: a nonergodic long-time behavior ($t \rightarrow \infty$) seen in the average occupation probability and captured by the combined enhancement factor of $P^{\textrm{RQB}}P^{\textrm{UQB}}$ entailing the memory of both the initial state and its early dynamics.} 
\label{fig:promo}
\end{figure}

As previously outlined in Sec.~II, QBs can be naturally expressed in terms of density matrices $\rho_a$ and $\rho_b$ [see Eq.~\eqref{Eq:quantum_phase_space_exploration_density_matrix}]. However, we have earlier set aside the possibility of impure density matrices with $\textrm{Tr}[\rho^2] < 1$. For instance, the probe state $\rho_a$ may represent an incoherent mixture of $n$, nearly nonoverlapping pure states, with each component given by $\rho_n = \sum_m a_m \vert E_m \rangle \langle E_m \vert$, where the $a_m$ are appropriate weights. Such a collection $\{\rho_j \mid j = 1, \dotsc, n \}$ may be regarded as covering an infinitesimal region of phase space. In the limit of large $n$, the density $\rho_a$ effectively is deprived of its quantum interference effects and memory. If the test states $\rho_b$ are constructed similarly, sharing matched energy envelopes with the density $\rho_a$, then in a quantum-ergodic system, $P_{aa} \approx P_{ab}$ for all such states, reflecting the underlying classical ergodicity purged from interference effects. Since the UQB factor acts as a proxy to quantum coherence, this path leads to $P^{\mathrm{UQB}} \sim 1$.

Nevertheless, the suppression of the universal component does not imply that the revival contribution is also lost. That is, the factor $P^{\mathrm{RQB}}$ may still deviate significantly from unity, shaped by the initial-decay profile of the nonstationary state. As the maximum exploration principle still holds, unless the system undergoes complete decoherence reducing to a fully fledged classical counterpart, a QB can form out of the early dynamics of the considered distribution occurring before the Heisenberg time.

Our QB work has antecedents in phase-space flow and quantum-ergodicity studies over several decades, but the explicit QB idea was first previewed in Ref.~\cite{graf_birthmarks_2024}. Subsequently and after discussions with us, a similar idea was previewed in Ref.~ \cite{pizzi_quantum_2024}. The later study~\cite{pizzi_phys.rev.lett_134_140402_2025}  examined an aspect similar to our QB but under the banner of ``quantum trails", basically reproducing our stadium results  in a simplified form. Despite acknowledging the existence of our prior QB theory discussed here, the work in question nevertheless fails to recognize several important points. It  inaccurately confines the phenomenon to narrowly peaked wave packets aligned with classical trajectories. As demonstrated here, any nonstationary state will cast a (universal) QB, and this effect is even further amplified by the recurrences and slow exploration rate (revival enhancement). This fact is not only proven here but we have also quantified it. Many of the open questions raised in Ref.~\cite{pizzi_phys.rev.lett_134_140402_2025}, such as the relevance of the phenomenon, are clearly addressed by the current work. In addition, the discussion in Ref.~\cite{pizzi_phys.rev.lett_134_140402_2025} totally overlooks a substantial body of existing literature thoroughly addressed and connected within our grand tapestry of QBs.

On another note, we emphasize that the QB phenomenon should not be confused with the Poincaré recurrence (PR)~\cite{Arnold_book}, which is purely classical in origin, nor with its loose quantum analog in finite Hilbert spaces, the full quantum revival (FQR)~\cite{styer_am.j.phys_69_56_2001, rasanen_eur_phys.j.b_86_17_2013}. First, both PR and FQR occur on astronomical time scales~\footnote{Even for integrable systems, this time can be tremendous, depending on the size of the Hilbert space, and it is even longer in chaotic ones. For both the soft- and hard-wall billiards considered here, thus timescale lies far beyond our simulation window and thus does not affect our analysis of QBs.}, vastly exceeding the characteristic QB scale, which is set by the Heisenberg time. Second, while PR and FQR concern only the reappearance of the initial state, a QB encodes not only the memory of the initial state but also of its full early-time dynamics, such as the influence of POs. Third, PR and FQR correspond to single, isolated events in time, whereas a QB manifests as a persistent statistical enhancement in the probability of revisiting the initial state and its subsequent evolutes.

Importantly, the QB phenomenon should not be misconstrued as a breakdown of the ETH~\cite{Srednicki_phys.rev.a_50_888_1994, Deutsch_phys.rev.a_43_2046_1991}, which remains a central framework for understanding quantum thermalization~\cite{Reichl_book}. Specifically, for a given initial state $\vert a \rangle$, ETH expresses the long-time average of an observable $\hat{\mathcal{O}}$ as a smooth mean with subleading fluctuations~\cite{Srednicki_phys.rev.a_50_888_1994, Deutsch_phys.rev.a_43_2046_1991, Reichl_book}, yielding $\bar{\mathcal{O}} = \sum_n p_n^{a} \mathcal{O}_{nn}$ in the limit $t \rightarrow \infty$. Crucially, the QB effect is not concealed in these vanishing fluctuations, but rather in the fine structure of the diagonal contributions themselves, namely, in the correlations between the matrix elements $\mathcal{O}_{nn} = \langle E_n\vert \hat{\mathcal{O}}\vert E_n \rangle$ and the expansion coefficients $p_n^{a} = \vert \langle E_n \vert a \rangle \vert^2$, defined here with respect to the eigenstates $\vert E_n \rangle$ of the Hamiltonian (cf. Eq.~\eqref{Eq:expansion_of_the_initial_state}). While the coefficients $p_n^{a}$ encode information about the initial state $\vert a \rangle$, the diagonal matrix elements $\mathcal{O}_{nn}$ themselves can also contain nontrivial correlations.

In the ETH framework, one typically assumes~\cite{Srednicki_phys.rev.a_50_888_1994, Deutsch_phys.rev.a_43_2046_1991, Srednicki_J.Phys.A_32_1163_1999, Rigol_Nature_452_854_2008, Rigol_Phys.Rev.Lett.108.110601_2012} that the elements $\mathcal{O}_{nn}$ vary smoothly, or are effectively constant, within a given energy window, therefore corresponding to an unbiased sampling over eigenstates $\vert E_n \rangle$. Under this assumption, the long-time expectation value $\bar{\mathcal{O}}$ coincides with the corresponding ensemble average. By contrast, our QB perspective makes no such additional assumptions, relying solely on the standard credo of quantum states in Hilbert space and their unitary time evolution. In fact, the QB effect stems from subtle correlations, which are commonly washed out in ETH-based reasoning. For example, choosing the operator $\hat{\mathcal{O}}$ as the identity operator directly leads to our QB enhancement of $\bar{P}_{aa}/\bar{P}_{ab} \ge 2$. Moreover, the QB concept extends beyond chaotic systems, even to encompass integrable ones (see, e.g., Ref.~\cite{Universal_quantum_birthmark_paper}), which lie outside the scope of conventional ETH~\cite{Rigol_Phys.Rev.Lett.108.110601_2012}.

The magnitude of the QB effect, nonetheless, depends on both the initial state $\vert a \rangle$ and the choice of observable $\hat{\mathcal{O}}$, the latter paralleling the circumstance of ETH. A systematic analysis of this dependence, i.e., how QB signatures appear in the expectation values of different classes of observables, constitutes an important direction for future work. From a broader perspective, we regard the QB concept as a complementary extension of the ETH framework: It reveals additional structure in quantum correlations that lies beyond the scope of the standard ETH ansatz, thereby providing a more nuanced understanding of thermalization from the shared standpoint of quantum dynamics and ergodicity.

\subsection{Connection to scarring and beyond}

While shedding light on the nebulous quantum nature of ergodicity, the QB phenomenon also introduces a generalization of the original insight behind the Heller-type scarring~\cite{Heller_phys.rev.lett_53_1515_1984}, extending its applicability to arbitrary nonstationary quantum states. This concept also extends the phase-space considerations of scarring by Bogomolny~\cite{bogomolny_physica.d_31_169_1988} and Berry~\cite{berry_proc_r_soc_lond_a_423_219_1989}. Moreover, the new framework applies to both variational~\cite{keski-rahkonen_phys.rev.lett_123_214101_2019} and many-body~\cite{serbyn_nat.phys_17_675_2021} scarring phenomena. 

Even though the former shares resemblance to the conventional scarring, it originates from a fundamentally distinct mechanism~\cite{keski-rahkonen_phys.rev.lett_123_214101_2019}: The (near) symmetries of the unperturbed system lead to scarred states favored by the variational principle in the presence of a localized perturbation. These variational scars form around the quasiperiodic orbits of the unperturbed classical counterpart, resulting in exceptionally strong wave-packet recurrences~\cite{luukko_sci.rep_6_37656_2016}. 

On the other hand, many-body scarring involves special initial states that exhibit long-lived, quasiperiodic revivals defying thermalization over experimentally accessible timescales, as observed in Rydberg atom arrays~\cite{bernien_nature_551_579_2017, turner_nat.phys_14_745_2018, Ho_phys.rev.lett_122_040603_2019, serbyn_nat.phys_17_675_2021} and optical lattices~\cite{scherg_nat.commun_12_1_2021, zhao_phys.rev.lett_124_160604_2020}. This ergodicity breaking arises from a small group of special eigenstates, i.e., many-body scars~\cite{serbyn_nat.phys_17_675_2021}, embedded in a sea of thermal eigenstates generically obeying the ETH. Shared among all forms of scarring, quantum recurrences or revivals prompt the existence of a RQB, which subsequently explains the failure of full ergodicity anticipated by the maximum rate principle.

The scope of QBs is broader than scars: QBs transcend the necessity of POs, applying to all types of trajectories. A wave packet launched along a classical trajectory will initially follow it closely, as required by the Ehrenfest theorem. This adherence consequently yields a moderately large factor of $P^{\textrm{RQB}}$, and thus there is a QB of this classical trajectory left in the long-term quantum behavior of the wave packet. Consistent with the maximum rate principle, this behavior results in the form of weak ergodicity breaking. It further follows that the spectrum must then contain eigenstates that support this dynamics~\cite{Antiscarring_1, lu_phys.rev.a_112_043307_2025}, retaining vestiges of the classical trajectory. Nonetheless, the manifestation of this effect can be particularly subtle, as shown by the variational scarring~\cite{keski-rahkonen_j.phys.conden.matter_31_105301_2019, keski-rahkonen_phys.rev.b_97_094204_2017, keski-rahkonen_phys.rev.lett_123_214101_2019} in the case of nearly degenerate eigenstates and corresponding classical POs.  

In addition to scarring, the analysis of the eigenstate properties plays a central role in understanding other quantum memory effects, such as semiclassical~\cite{kaplan_phys.rev.e_62_409_2000, bohigas_phys.rep_223_43_1993}, Anderson~\cite{anderson_phys.rev_109_1492_1958, Anderson_localization_book, Fartash_Phys.Rev.B_113_195401_2026}, dynamical~\cite{casati_dynamical_localization, germpel_phys.rev.A_29_1639, shepelyansky_physica.d_28_103_1987, izrailev_phys.rep_196__299_1990}, and many-body~\cite{pal_phys.rev.b_82_174411_2010, abanin_rev.mod.phys_91_021001_2019} localization. Furthermore, our analysis of QBs directly extends to quantum maps~\cite{berry_ann.phys_122_26_1976, oconnor_ann.phys_207_218_1991} with discrete time evolution. Systems such as the kicked rotor~\cite{Stockmann_book} and the quantum Baker map~\cite{oconnor_ann.phys_207_218_1991} present an interesting avenue for future research. 

\subsection{Application and experiment}\label{Discussion_C}

The phenomenon likely plays a role in systems with mixed phase-space structure~\cite{backer_j.phys.a_35_527_2002} and in the statistical behavior of resonance wavefunctions in open quantum systems~\cite{Lu_phys.rev.lett_82_5233_1999}. Applications of these kind of chaotic Hamiltonians range general quantum-dot physics~\cite{Nakamura_book, Schoeppl_Phys.Rev.B_113_165402_2026, Lin_Proc.Natl.Acad.Sci_123_e2538081123_2026}, like fractal conductance~\cite{crook_phys.rev.lett_91_246803_2003,Ketzmerick_phys.rev.b_54_10841_1996, Sachrajda_phys.rev.lett_80_1948_1998} and transport in the Coulomb blockade regime~\cite{kaplan_phys.rev.b_78_085305_2008, tomsovic_phys.rev.lett_100_164101_2008, jalabert_phys.rev.lett_65_2442_1990}, to the control of directional emission in microcavity lasers~\cite{nockel_nature_385_45_1997, gmachl_science_280_1556_1998, tureci_phys.rev.a_74_043822_2006, chern_appl.phys.lett_83_1710_2003, rex_phys.rev.lett_88_094102_2002}. For example, in ballistic quantum dots containing even thousands of electrons, several statistical measures commonly exceed the random wave predictions by factors of 3 or more~\cite{kaplan_phys.rev.b_78_085305_2008, tomsovic_phys.rev.lett_100_164101_2008}, indicating the presence of a QB.

To further underline the relevance of  QB, we draw attention to two additional cases. First, we have identified the UQB in the twin-peak formation~\cite{akkermans_phys.rev.lett_56_14711986, karpiuk_phys.rev.lett_109_190601_2012, jendrzejewski_phys.rev.lett_109_195302_2012, cherroret_phys.rev.a_85_011604_2012, micklitz_phys.rev.lett_112_110602_2014, lee_phys.rev.a_90_043605_2014, ghosh_phys.rev.a_90_063602_2014, gosh_phys.rev.a_95_041602_2017}; because of Anderson localization, a coherent backward-scattering peak is eventually accompanied by an interference peak in the forward direction of the momentum distribution when a quantum wave packet is launched with a finite momentum into a disordered potential landscape. In the long-time limit, the forward- and backward-scattering peaks are identical but more interestingly the peak height is thrice the overall scattering as predicted by a UQB. 
In the coordinate space, this behavior leads to the so-called quantum boomerang effect~\cite{prat_phys.rev.a_99_023629_2019}.

Second, the RQB effect and the maximum rate principle appear in the electronic absorption spectra of medium-sized molecules at moderate energies. For instance, the vibrational spectrum of the acetylenic stretch in propynol~\cite{hudspeth_j.chem.phys_109_4316_1998}, when purified from thermal and rotational background via double resonance methods, consists of extremely narrow bands. A single transition line broadens into a narrow band caused by anharmonic interactions, but the observed band is surrounded by wide spectral regions devoid of intensity. Without prior knowledge of the constrained, slow initial escape, this desert would appear as a dramatic failure of ergodicity.
Of course, if the narrow bands are acknowledged as a prior constraint, the system can be regarded as nearly ergodic within that restricted domain.  

On top of the two emphasized examples above, we want to point out an interesting possibility with modern quantum simulators, such as in Refs.~\cite{bernien_nature_551_579_2017, turner_nat.phys_14_745_2018, Ho_phys.rev.lett_122_040603_2019, scherg_nat.commun_12_1_2021, zhao_phys.rev.lett_124_160604_2020} that enables one to investigate QBs in both quantum maps and many-body scarring contexts, with the distinct advantage of controllable Hilbert-space dimensionality. Along the same line, there is the relationship between QBs and decoherence~\cite{gorin_jopt.b_4_s386_2002, gorin_phys.lett.a_309_61_2003, gorin_phys.rep_435_33_2006, pineda_phys.rev.a_75_012106_2007}, particularly in the context of open quantum systems interacting with their environment~\cite{gelbart_j.chem.phys_57_4699_1972, mulhall_rev.phys.c_91_014305_2015, beenakker_rev.mod.phys_69_731_1997}. It may be possible to relate QBs to the purity decay of one or two qubits~\cite{pineda_new.j.phys_9_106_2007,cerruti_j.phys.a_36_3451_2003, gorin_new.j.phys_6_20_2004, frahm_eur.phys.j.d_29_139_2004}. Moreover, as commonly investigated in the milieu of quantum simulators, a state $\vert a \rangle$ can start in a sticky Hilbert-space region, such as linked to many-body scars~\cite{serbyn_nat.phys_17_675_2021} or a fragmented part of Hilbert space~\cite{sala_phys.rev.x_10_011047_2020}, which can effectively lead to the QB enhancement of applying to the whole special subspace compared to the nominal Hilbert space, paving way to an extension of QBs, namely, {\it birthplaces}.

\subsection{Birthplaces}

Generally speaking, QBs are regions of phase space visited by an evolving state $\vert a\rangle$; if its evolute $\vert \alpha \rangle$ becomes momentarily concentrated, it leads to a persistent enhancement in the long-time visitation probability. This raises the question of whether such enhancement can extend to affect entire zones, effectively forming an enhanced birthplace. This constitutes a violation of classical ergodicity: The entire birthplace is then statistically remembered.

A representative coordinate-space scenario considers a wave packet launched in a small chamber connected to a much larger one through a narrow channel acting as a bottleneck, with both regions exhibiting classically chaotic dynamics. In such a setup, the long-time probability distribution may remain biased toward the smaller chamber where the wavefunction was initially localized. The underlying mechanism is that the wave packet experiences a delay while searching for escape paths from the small into the larger chamber. During this time, dynamical recurrences, even occurring on timescales up to the Heisenberg time of the isolated small chamber, impede the uniform exploration of phase space. Consequently, the amplitude of the wave packet escaping the small room must abide by this slower rate, which will leave the larger chamber less than fully explored by the Heisenberg time of the whole system. If the escape is very slow from the first chamber, such as having to tunnel out of it, the effect could be very dramatic. 

In fact, this principle has effectively been seen in the context of quantum transport across classical partial barriers modeled by block-structured random matrices with weakly coupled subspaces, as previously demonstrated in Ref.~\cite{michler_phys.rev.lett_109_234101_2012}. Therefore, the concept of birthplaces couples the QB phenomenon to the less-studied dynamics induced by block-structure (chaotic) Hamiltonians, which generalize the core idea of RMT and open a possible new avenue for the future QB research.

\section{Summary} \label{Sec:conclusion}

In summary, there is a more intricate and nuanced structure of nonergodic wave behavior in classically chaotic systems than previously recognized. A QB emerges from any nonstationary state in an isolated quantum system governed by unitary evolution. This notion of a QB surpasses the analysis of individual eigenstates, as well as reestablishing the lost connection to the classical, dynamics-based standpoint of ergodicity. Instead of treating universal RMT and system-specific short-time behavior separately, we have shown that they can be naturally come together in a QB. Particularly, the existence of QBs present a salient, dynamical correction onto the blank adage of quantum ergodicity and thermalization.

\section*{Acknowledgments}

The authors thank L. Kaplan for many inspiring discussions. We also thank S. Yelin, A. Aydin and C. Dag for valuable feedback. This work was supported by the National Science Foundation under Grant No. 2403491. A.M.G. thanks the Studienstiftung des Deutschen Volkes for financial support. R.K. was funded by the Deutsche Forschungsgemeinschaft (DFG, German Research Foundation) Grant No. 262765445. J.K.-R. thanks the Oskar Huttunen Foundation for the financial support.

\bibliography{references}
\clearpage  
\appendix

\section{Derivation of the universal quantum birthmark} \label{Appx:UQB_derivation}

This appendix provides a simplified mathematical derivation of the universal quantum birthmark (UQB), a persistent memory effect in the long-time dynamics of (fully) chaotic quantum systems. We first establish the statistical framework and the ergodic baseline. We then derive the universal enhancement of the long-time return probability employing the $\chi^2$ distribution of spectral projections. For completeness, we also present a symmetry-based derivation that treats normalization and finite-$N$ correlations exactly. Finally, we show that this enhancement is invariant under unitary time evolution, constituting a ``birthmark'' imprinted on a quantum state and preserved throughout its subsequent evolution. An additional proof, including a rigorous derivation of the universal UQB factors of 2 (or 3) and a discussion of how additional symmetries modify these factors, is provided in Ref.~\cite{Universal_quantum_birthmark_paper}.

\subsection{Preliminaries and mathematical setup}

We consider a generic quantum system described by a Hamiltonian $\hamil$ acting on an $N$-dimensional Hilbert space $\mathcal{H}$. The system is assumed to be chaotic in the classical limit, with its statistical properties described by random matrix theory (RMT). Specifically, our framework entails the following aspects:

\begin{itemize}
    \item \textbf{Eigenstates and energies:} The Hamiltonian has a complete, orthonormal set of energy eigenstates $\{\ket{E_n}\}$ with corresponding non-degenerate energies $\{E_n\}$, where $n = 1, \dots, N$.
    
    \item \textbf{Non-stationary states:} We consider two generic, nonstationary states, $\ket{a}$ and $\ket{b}$, which are normalized vectors in $\mathcal{H}$. These states are typically localized, e.g., Gaussian wave packets.
    
    \item \textbf{Spectral projections:} A nonstationary state $\ket{a}$ can be expanded in the energy eigenbasis as
    \begin{equation}
        \ket{a} = \sum_{n=1}^{N} a_n \ket{E_n},
    \end{equation}
    where $a_n = \langle E_n | a \rangle$. The probability of finding the system in the eigenstate $\ket{E_n}$ when it is in state $\ket{a}$ is given by
    \begin{equation}
        p_n^a = |a_n|^2 = |\langle E_n | a \rangle|^2,
    \end{equation}
    with normalization $\sum_{n=1}^{N} p_n^a = 1$.
    
    \item \textbf{Long-time average transition probability:}The long-time average probability of finding a system initially in state $\ket{a}$ in state $\ket{b}$ is given by their spectral overlap. We denote this quantity by $P_{ab}$, defined as
    \begin{equation}
    \begin{split}
        P_{ab} &= \lim_{T\to\infty} \frac{1}{T} \int_0^T |\bra{b}e^{-i\hamil t/\hbar}\ket{a}|^2 dt\\
        &= \sum_{n=1}^{N} p_n^a p_n^b.
    \end{split}
    \end{equation}
    The cross terms in the time integral vanish upon averaging because the energy spectrum is nondegenerate.
    
    \item \textbf{Self-overlap (dilution factor):} For $\ket{b} = \ket{a}$, the transition probability reduces to the self-overlap, also known as the dilution factor or the inverse participation ratio (IPR) in the energy basis. It is the long-time average probability of returning to the initial state, formally
    \begin{equation}\label{Appx:Eq_dilation_factor}
        P_{aa} = \sum_{n=1}^{N} (p_n^a)^2.
    \end{equation}
\end{itemize}

\subsection{The random-matrix-theory baseline and the universal enhancement of self-overlap} \label{App:UQBproof}

In a fully ergodic system, all accessible states are sampled with equal probability. In the present setting, this establishes a baseline for the transition probability $P_{ab}$ between two distinct states, $\ket{a}$ and $\ket{b}$. We derive this baseline and the corresponding UQB enhancement in two ways: (I) a probabilistic argument based on statistical independence, and (II) a symmetry-based derivation that incorporates exact normalization and finite-$N$ correlations.

\subsubsection*{(I) Probabilistic proof via the $\chi^2$ distribution}

For pedagogical clarity, we first consider a probabilistic derivation in which the expansion coefficients are modeled as independent Gaussian variables. This provides a simple representation of the universal RMT statistics. In a chaotic system, the eigenstates behave as pseudo-random vectors, and the spectral projections of two distinct generic states are statistically uncorrelated. The resulting independence assumption provides an appropriate RMT approximation for the interaction between two unrelated states and reproduces the correct large-$N$ behavior. Although it does not impose the normalization constraint exactly for finite $N$, the resulting corrections vanish in the large-$N$ limit. It therefore provides a convenient reference for quantifying the nonergodic enhancement.

We adopt the following RMT assumptions:
\begin{enumerate}
    \item For any generic state $\ket{x}$, the ensemble-averaged projection probability is uniform across the spectrum, i.e., $\expval{p_n^x} = 1/N$.
    This value follows directly from the normalization condition, $\sum_n \expval{p_n^x} = N \expval{p_n^x} = 1$.

    \item For two distinct and uncorrelated states $\ket{a}$ and $\ket{b}$, the spectral projections $p_n^a$ and $p_n^b$ are statistically independent, i.e., $\expval{p_n^a p_n^b} = \expval{p_n^a}\expval{p_n^b}$.
\end{enumerate}
Under these standard assumptions, the ensemble average of $P_{ab}$ is
\begin{equation}
\begin{split}
    \expval{P_{ab}} &= \expval{\sum_{n=1}^{N} p_n^a p_n^b}\\ 
    &= \sum_{n=1}^{N} \expval{p_n^a p_n^b} = \sum_{n=1}^{N} \expval{p_n^a}\expval{p_n^b}.
\end{split}
\end{equation}
Consequently, substituting the uniform average probability gives
\begin{equation}\label{eq:baseline}
    \expval{P_{ab}} = \sum_{n=1}^{N} \left(\frac{1}{N}\right) \left(\frac{1}{N}\right) = N \left(\frac{1}{N^2}\right) = \frac{1}{N}.
\end{equation}
Thus, we see that $\expval{P_{ab}} = 1/N$ defines the ergodic baseline. A fully ergodic system therefore yields a transition probability of $1/N$ for a typical target state.

The self-overlap $P_{aa}$, however, differs from the ergodic baseline because of fluctuations in the spectral projection probabilities $p_n^a$, and its ensemble average is therefore enhanced. We define the corresponding (universal) enhancement factor as
\begin{equation}
    P^{\textrm{UQB}} = \frac{\expval{P_{aa}}}{\expval{P_{ab}}} = \frac{\expval{P_{aa}}}{1/N} = N \expval{P_{aa}}.
\end{equation}
The evaluation of $\eta$ reduces to the second moment of the distribution of $p_n^a$. Since $\expval{P_{aa}} = \expval{\sum_n (p_n^a)^2} = N \expval{(p_n^a)^2}$, we end up with $\eta = N^2 \expval{(p_n^a)^2}$.

Within the RMT framework established above, the projection probabilities $p_n^a$ follow a scaled $\chi^2$ distribution. This fact follows from the Gaussian statistics of the coefficients $a_n$, which are either real or complex valued depending on the presence or absence of time-reversal symmetry, respectively. In other words, we can write
\begin{equation}
    p_n^a \sim \frac{1}{N\beta} \chi^2_\beta,
\end{equation}
where $\beta$ is the Dyson index: $\beta=1$ for systems with time-reversal symmetry (GOE statistics) and $\beta=2$ for systems without time-reversal symmetry (GUE statistics).

The moments of a standard $\chi^2_\beta$ distribution are the first moment (mean) $E[\chi^2_\beta] = \beta$ and second moment (variance) of the distribution $\text{Var}(\chi^2_\beta) = 2\beta$ of the distribution. Therefore, the mean and variance of $p_n^a$ are
\begin{equation}
\begin{split}
    \expval{p_n^a}
    &= \mathbb{E}\left[\frac{1}{N\beta} \chi^2_\beta\right]\\
    &= \frac{1}{N\beta} \mathbb{E}[\chi^2_\beta]\\
    &= \frac{1}{N\beta}\beta\
    = \frac{1}{N}
\end{split}
\end{equation}
and
\begin{equation}
\begin{split}
    \text{Var}(p_n^a)
    &= \text{Var}\left(\frac{1}{N\beta} \chi^2_\beta\right)\\
    &= \left(\frac{1}{N\beta}\right)^2 \text{Var}(\chi^2_\beta)\\
    &= \frac{1}{N^2\beta^2}2\beta
    = \frac{2}{N^2\beta}.
\end{split}
\end{equation}
Subsequently, we see that
\begin{equation}
\begin{split}
    \expval{(p_n^a)^2}
    &= \text{Var}(p_n^a) + (\expval{p_n^a})^2\\
    &= \frac{2}{N^2\beta} + \left(\frac{1}{N}\right)^2\\
    &=\frac{1}{N^2}\left(1 + \frac{2}{\beta}\right),
\end{split}    
\end{equation}
which yields the average self-overlap
\begin{equation}
\begin{split}
    \expval{P_{aa}}
    &= N \expval{(p_n^a)^2}\\
    &= N \cdot \frac{1}{N^2}\left(1 + \frac{2}{\beta}\right)\\
    &= \frac{1}{N}\left(1 + \frac{2}{\beta}\right).
\end{split}  
\end{equation}
We can thus conclude the enhancement factor to be
\begin{equation}
    P^{\textrm{UQB}} = N \expval{P_{aa}} = 1 + \frac{2}{\beta},
\end{equation}
or more specifically
\begin{equation}
    P^{\textrm{UQB}} = 
    \begin{cases}
        2 & \textrm{when}\, \beta = 2\, (\textrm{GUE}),\\
        \\
        3 & \textrm{when}\, \beta = 1\, (\textrm{GOE}).\\
    \end{cases}
\end{equation}
In other words, the universal enhancement factor is $P^{\textrm{UQB}} = 2$ or $P^{\textrm{UQB}} = 3$, depending on whether time-reversal symmetry is absent or present in the initial state and quantum-chaotic system that likewise determines whether the expansion coefficients are real or complex.

The probabilistic-based derivation above reproduces the correct large-$N$ limits but does not enforce exact normalization or account for finite-$N$ correlations. We therefore turn to a symmetry-based derivation in the next subsection that treats these constraints exactly.

\subsubsection*{(II) Symmetry-based derivation: Fourth moments from unitary and orthogonal invariance}

We next derive the result of the universal enhancement factor from unitary and orthogonal invariance. Unlike the preceding $\chi^2$-based argument, which treats the components as independent and imposes normalization only on average, this approach enforces normalization exactly. The relevant fourth moments are then fixed by symmetry, making the universality of the result and its finite-$N$ corrections explicit.

We start by consider a normalized random vector $\mathbf{a} = (a_1,\dots,a_N)$ composed of the expansion coefficients $a_n$ ($n = 1, \dots, N$) of a generic initial state $|a\rangle$. In a chaotic system, this vector is thus distributed uniformly on the unit sphere in an $N$-dimensional Hilbert space. The distribution is invariant under unitary transformations for complex vectors (GUE case) and under orthogonal transformations for real vectors (GOE case). In particular, we are interested in the second moment of the spectral probabilities $p_n^a = |a_n|^2 $, which determines the ensemble average of the self-overlap $P_{aa} = \sum_{n=1}^N (p_n^a)^2$ [see Eq.~\eqref{Appx:Eq_dilation_factor}].

In the GUE case where the vector $\mathbf{a}$ is complex, we define the fourth-order moment tensor
\begin{equation}
T_{\alpha\beta\gamma\delta}
= \mathbb{E}\!\left[ c_\alpha c_\beta^{*} c_\gamma c_\delta^{*} \right],
\end{equation}
where the indices $\alpha, \beta, \gamma, \delta \in \{ 1, \dots, N \}$. Because the vector $\mathbf{a}$ is invariant under a unitary transformation $U(N)$, the tensor $T_{\alpha\beta\gamma\delta}$ must also be invariant under simultaneous transformations of all indices. By the Schur-Weyl duality, the space of such invariant tensors is spanned by pairings between vector and dual-vector indices. Thus, the most general fourth-order invariant tensor can be decomposed as
\begin{equation}
T_{\alpha\beta\gamma\delta}
= A\,\delta_{\alpha\beta}\delta_{\gamma\delta}
+ B\,\delta_{\alpha\delta}\delta_{\beta\gamma},
\label{eq:GUEtensor}
\end{equation}
with coefficients $A$ and $B$ independent of the indices.

We now fix $A$ and $B$ using normalization, for which we define $C \equiv \mathbb{E}[|c_j|^4]$ and $D \equiv \mathbb{E}[|a_i|^2 |a_j|^2]$ with $i\neq j$. Setting $\alpha=\beta\neq\gamma=\delta$ in the tensor of Eq.~\eqref{eq:GUEtensor} gives $D=A$, while $\alpha=\delta\neq\beta=\gamma$ gives $D=B$; together these imply $A=B$. On the other hand, taking all indices equal gives $C = T_{iiii} = 2A$, so that $D=C/2$. Imposing the normalization condition $\mathbb{E} \left[\left(\sum_{j=1}^N |a_j|^2 \right)^2 \right] = 1$ gives $NC + N(N-1)D = 1$, and substituting $D=C/2$ leads to
\begin{equation}
NC + \frac{N(N-1)}{2}C = 1
\, \Longrightarrow \;
C = \frac{2}{N(N+1)},
\end{equation}
which subsequently means $\mathbb{E}[(p_n^a)^2] = \mathbb{E}[|a_i|^4] = 2/[N(N+1)]$.

We can proceed similarly in the case of GOE as above, but the vector $\mathbf{a}$ is then real valued and the relevant object is the fourth-order tensor $T_{ijkl} = \mathbb{E}[a_i a_j a_k a_l]$. Orthogonal invariance further restricts it to the form
\begin{equation}\label{eq:GOEtensor}
T_{ijkl}
= X\,\delta_{ij}\delta_{kl}
+ Y\,\delta_{ik}\delta_{jl}
+ Z\,\delta_{il}\delta_{jk}.
\end{equation}
We again define $C=\mathbb{E}[a_j^4]$ and $D=\mathbb{E}[a_i^2 a_j^2]$ for $i\neq j$. Evaluating the tensor in Eq.~\eqref{eq:GOEtensor} at $(i,i,j,j)$ permutations shows $X=Y=Z=D$, while taking all indices equal gives $C=T_{iiii}=X+Y+Z=3D$. Then, the normalization condition $\mathbb{E} \left[ \left(\sum_{j=1}^N a_j^2 \right)^2 \right] = 1$ gives $NC+N(N-1)D=1$. When using $D=C/3$, we get
\begin{equation}
NC + \frac{N(N-1)}{3}C = 1
\, \Longrightarrow \,
C = \frac{3}{N(N+2)}.
\end{equation}
In other words, we see that $\mathbb{E}[(p_n^a)^2] = \mathbb{E}[a_i^4] = 3/[N(N+2)]$ for GOE.

Collectively, we obtain
\begin{equation}
\langle P_{aa} \rangle
=
\begin{cases}
\dfrac{2}{N+1}, & \text{GUE},\\[6pt]
\dfrac{3}{N+2}, & \text{GOE},
\end{cases}
\end{equation}
by recalling $P_{aa} = \sum_{n=1}^N p_n^2$. Furthermore, since the ensemble average satisfies $\langle P_{ab} \rangle = 1/N$ for two independent states $|a\rangle$ and $|b\rangle$, the universal enhancement factor is 
\begin{equation}
P^{\textrm{UQB}} =
\frac{\langle P_{aa} \rangle}{\langle P_{ab} \rangle}
=
\begin{cases}
\dfrac{2N}{N+1}, & \text{GUE},\\[6pt]
\dfrac{3N}{N+2}, & \text{GOE}.
\end{cases}
\end{equation}
Notably, in the limit of $N\gg1$, these factors approach the earlier, heuristic bounds of 2 and 3 for GUE and GOE, respectively.

\subsection{Physical interpretation and the persistent birthmark}

Generally speaking, the self-overlap $P_{aa}$ is the inverse participation ratio (IPR) of the state $\ket{a}$ in the energy eigenbasis, measuring the inverse effective number of eigenstates contributing to the state $\ket{a}$. It is also the infinite-time-averaged return probability, quantifying the long-time likelihood of returning to the initial state. Our result $\expval{P_{aa}} \approx P^{\textrm{UQB}}/N$ therefore shows that, on average, a chaotic state has a higher return probability than predicted by a simple ergodic argument.

We next show that this enhanced self-overlap is not restricted to the initial state $\ket{a}$, but persists throughout its entire evolution. Let $\ket{\alpha}$ denote the state propagated of the initial state $| a \rangle$ for time $t$, i.e.,
\begin{equation}
\ket{\alpha} = e^{-i\hamil t/\hbar}\ket{a}.
\end{equation}
Its spectral projections are $p_n^\alpha = |\langle E_n | \alpha \rangle|^2$, with
\begin{equation}
\langle E_n | \alpha \rangle = \bra{E_n} e^{-i\hamil t/\hbar} \ket{a}.
\end{equation}
Since $\bra{E_n}\hamil = E_n\bra{E_n}$, we have $\bra{E_n} e^{-i\hamil t/\hbar} = e^{-iE_n t/\hbar} \bra{E_n}$, meaning that
\begin{equation}
\langle E_n | \alpha \rangle = e^{-iE_n t/\hbar} \langle E_n | a \rangle.
\end{equation}
Combining these results above, we obtain that
\begin{equation}
\begin{split}
p_n^\alpha &= |e^{-iE_n t/\hbar} \langle E_n | a \rangle|^2\\ 
&= |e^{-iE_n t/\hbar}|^2 |\langle E_n | a \rangle|^2
= p_n^a
\end{split}
\end{equation}
since $|e^{-iE_n t/\hbar}|^2 = 1$ for any energy $E_n$ at any time $t$, and $p_n^a = |\langle E_n | a \rangle|^2$. In particular, we see that
\begin{equation}
p_n^\alpha = p_n^a \quad \text{for all } n \text{ and } t.
\end{equation}
In other words, the probability distribution over the energy eigenstates is invariant under time evolution: although $\ket{\alpha}$ evolves in Hilbert space, its projections onto the energy eigenbasis remain unchanged.

Consequently, the self-overlap of the evolved state is time-independent:
\begin{equation}
P_{\alpha\alpha}(t) = \sum_{n=1}^{N} (p_n^\alpha)^2 = \sum_{n=1}^{N} (p_n^a)^2 = P_{aa},
\end{equation}
which immediately gives
\begin{equation}
\expval{P_{\alpha\alpha}} = \expval{P_{aa}} = \frac{P^{\textrm{UQB}}}{N}.
\end{equation}
Every state $\ket{\alpha}$ along the dynamics of the state $\ket{a}$ therefore retains the same enhanced self-overlap. This permanent, time-invariant signature, determined by the initial state and the system's symmetries, constitutes the UQB studied in the main text.

\section{Wave-packet simulations of quantum birthmarks} \label{Appx:Wavepacket_simulations}

\subsection{The split-operator method}

The time evolution of a quantum wave packet in the soft Bunimovich stadium potential can be efficiently simulated using the half-step split-operator method (Strang splitting). This approach yields a third-order accuracy in the time step, with a global error scaling as $\mathcal{O}(\Delta t^3)$, where $\Delta t$ denotes the temporal step size. To propagate the wavefunction $\psi(x, t)$ according to the time-dependent Schr{\"o}dinger equation, the Hamiltonian is decomposed into the kinetic energy operator $T = T(\mathbf{p})$ and the potential energy operator $V = V(\mathbf{r})$. The propagation is performed by first applying the potential operator in coordinate space for a half-time step, followed by a Fourier transform $\mathcal{F}$ to momentum space, where the kinetic operator is applied. An inverse Fourier transform $\mathcal{F}^{-1}$ then returns the wavefunction to coordinate space, where the second half of the potential operator is applied. This sequence yields the following expression for a full time step from $t$ to $t + \Delta t$, formally
\begin{equation}
\begin{split}
\psi(\mathbf{r}, t + \Delta t) &= e^{-i V(\mathbf{r}) \Delta t / 2} \times \\ &\mathcal{F}^{-1} \Bigg\{ e^{-i T(\mathbf{p}) \Delta t} \\ &\times \mathcal{F} \left\{ e^{-i V(\mathbf{r}) \Delta t / 2} \, \psi(\mathbf{r}, t) \right\} \Bigg\},
\end{split}
\end{equation}
where $\mathbf{r}$ and $\mathbf{p}$ denote the position and momentum coordinates, respectively.

\subsection{System setup}

For the wave-packet simulations, position and momentum space are discretized on $1024 \times 1024$ square grids. The coordinate-space grid spans $x_{\min} = -2.5 \times 10^{-8} \, \mathrm{m}$ to $x_{\max} = 2.5 \times 10^{-8} \, \mathrm{m}$, with grid spacing $\Delta x = (x_{\max} - x_{\min}) / 1024$ and $\Delta y = \Delta x$. The corresponding momentum-space grid consists of uniformly spaced frequency components centered at zero, with spacing $dk = 2\pi/L_{\rm box}$, where $L_{\rm box} = x_{\max} - x_{\min} = 5 \times 10^{-8} \, \mathrm{m}$ denotes the numerical box size. A time step of $\Delta t = 0.1 \, \textrm{fs}$ is used throughout.

The soft stadium potential is implemented analytically as
\begin{equation}
V(x, y) = \frac{V_0}{1 + \lambda e^{\mu [1 - f(x,y)]}},
\end{equation}
where the geometry and softness of the boundary are encoded in the dimensionless function
\begin{equation}
\begin{split}
f(x,y)  &=
\frac{(|x| - \frac{1}{2} L_{\rm s})^{2}}{R^{2}}
\Big[ H\left(x - \frac{1}{2} L_{\rm s}\right)\\ &+ H\left(-x - \frac{1}{2} L_{\rm s}\right) \Big]+\frac{y^{2}}{R^{2}},
\end{split}
\end{equation}
where $H(\cdot)$ is the Heaviside function. Here, $V_0 = 70 \, \mathrm{eV}$ is the maximum potential height, $R = 8.75 \times 10^{-9} \, \mathrm{m}$ is the radius of the semicircular end caps, and $L_{\rm s} = 1.5 \times 10^{-8}\,\mathrm{m}$ is the length of the stadium between them. The parameters $\lambda = 25$ and $\mu = 5$ control the softness and effective range of the potential wall, respectively.

The initial states $| a \rangle$ are Gaussian wave packets of the form
\begin{equation}
\begin{split}
\Psi(\mathbf{r}, 0) &= \langle \boldsymbol{r} \vert a\rangle\\ 
&= Z\exp\left[ -\frac{1}{4} \vert (\mathbf{r} - \mathbf{r}_0 ) \cdot \boldsymbol{\sigma} \vert^2 + i \mathbf{k} \cdot \mathbf{r} \right],
\end{split}
\end{equation}
where $\mathbf{r}_0 = (x_0, y_0)$ specifies the initial wave-packet center and $\mathbf{k}$ determines its initial momentum, and $Z$ is a normalization constant determined by the simulation box. For cases (A)--(C), the wave packet is centered at $\mathbf{r}_0=(0,0)$, whereas for case (D) it is initialized at $\mathbf{r}_0=(7.5 \times 10^{-9} \, \mathrm{m},,4.3 \times 10^{-9} \, \mathrm{m})$ to probe more general dynamics. In all cases, $\boldsymbol{\sigma} = (\sigma_x^{-1}, \sigma_y^{-1})$, with $\sigma_x=\sigma_y=2.5\times10^{-10} \, \mathrm{m}$, and the wavevector magnitude is fixed at $|\mathbf{k}| = (\pi / 2) \times 10^{10} \, \mathrm{m}^{-1}$. Taking the horizontal and vertical stadium axes as the $x$ and $y$ axes, respectively, the wavevector is oriented at $90^\circ$, $0^\circ$, $57^\circ$, and $123^\circ$ for cases (A), (B), (C), and (D), respectively, measured counterclockwise from the positive $x$ axis.

\subsection{Visualization of quantum birthmarks}

To visualize quantum birthmarks in coordinate space, we consider the time-averaged probability density $\bar{Q}(\mathbf{r})$, which characterizes local deviations from ergodicity:
\begin{equation}
\begin{split}
\bar{Q}(\mathbf{r}) &= A \lim_{T_{\max} \rightarrow \infty} \frac{1}{T_{\max}} \int_{0}^{T_{\max}} |\Psi(x,y,t)|^{2}, dt\\
&\longrightarrow 
\frac{A}{T_{\max}-T_{0}} \sum_{t = T_{0}}^{T_{\max}} |\Psi(x,y,t)|^{2}, \Delta t,
\end{split}
\end{equation}
where $\Psi(x,y,t)$ is the time-evolved wave packet and $A$ is the area of the billiard. The normalization is chosen such that an ergodic distribution gives $\bar{Q}(\mathbf{r})=1$, with values above and below unity indicating local enhancement and suppression, respectively. We refer to $\bar{Q}(\mathbf{r})$ as the scaled time-averaged probability density. For the soft-wall stadium, $A$ is taken to be the classically enclosed area at the mean energy of the wave packet.

Convergence was verified by varying $T_{0}$ while adjusting $T_{\max}$ accordingly, yielding consistent spatial enhancements. The resulting $\bar{Q}(x,y)$ for Cases (A)--(D) is shown in the main text. For a more detailed comparison of their spatial structure, Fig.~\ref{fig:1DCuts} shows one-dimensional cross-sections along the three representative lines indicated by the dashed lines in the corresponding panels.

\begin{figure*}[t]
\centering
\includegraphics[width=0.8\textwidth]{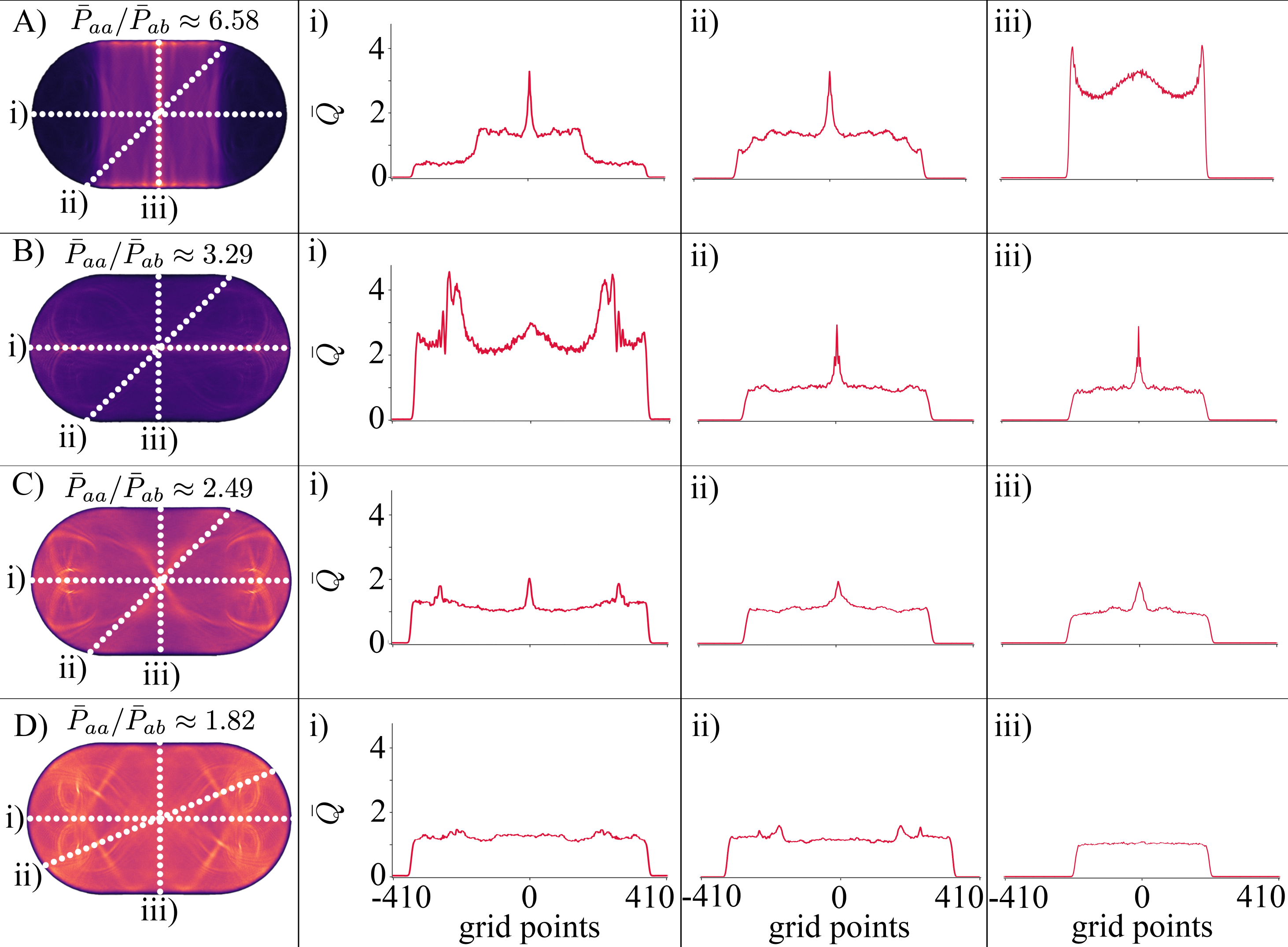}
\caption{One-dimensional cross-sections of the scaled time-averaged probability density $\bar{Q}(x,y)$ for Cases (A)--(D). Columns (i)--(iii) correspond to the three representative cross-section lines indicated by the dashed lines in the corresponding panels.}
\label{fig:1DCuts}
\end{figure*}

\subsection{Computation of the autocorrelation and fidelity}

To characterize the coherence and recurrence of the quantum wave packet, we compute the autocorrelation function $P(t)$, which quantifies the fidelity between the evolved state at time $t$ and the initial state:
\begin{equation}
\begin{split}
P(t)&=\left|\langle\Psi(0)|\Psi(t)\rangle \right|^2 \\ &\longrightarrow \left|\sum_{x,y}\Psi(x,y,0)^*\Psi(x,y,\tau)\Delta x \Delta y \right|^2
\end{split}
\end{equation}
where $\Psi(t)=\Psi(x,y,t)$ denotes the time-evolved wave packet, $\langle\Psi(0)|\Psi(t)\rangle$ is the spatial inner product, and $\Delta x$ and $\Delta y$ are the grid spacings along the $x$ and $y$ directions, respectively.

\subsection{Estimation of the Heisenberg time}

To assess the temporal structure of the quantum dynamics in the soft Bunimovich stadium, we express all time-dependent results in units of the Heisenberg time $t_H$. This is the characteristic timescale associated with resolving the discrete energy spectrum and provides a natural temporal scale for the dynamics, closely related to the maximum rate principle discussed in the main text. We estimate $t_H$ from the mean energy-level spacing $\Delta E$ within the energy window sampled by the wave packet. Since all wave packets have the same mean energy, a common value of $t_H$ is employed for all cases:
\begin{equation}
t_H \approx \frac{\hbar}{\Delta E},
\label{eq}
\end{equation}
where $\hbar$ is the reduced Planck constant. The mean level spacing is estimated from the density of states $g(E)$ as
\begin{equation}
\Delta E \approx \frac{1}{g(E)}.
\label{eq}
\end{equation}

For a single particle in a two-dimensional billiard, Weyl's law gives
\begin{equation}
g(E) \approx A \frac{m_e}{2\pi\hbar^2},
\label{eq}
\end{equation}
where $A$ is the classically accessible area and $m_e$ is the electron mass. Although this expression is strictly derived for hard-wall systems with stationary eigenstates, we utilize it here as an approximation for the soft-wall potential and nonstationary Gaussian wave packets. We determine $A$ from the area enclosed by the equipotential contour $V(x,y)=E$, with $E=\hbar^2 k^2/2m_e$ corresponding to the mean kinetic energy of the wave packet and $k=|\mathbf{k}|$. This gives $A=6.019\times10^{-16}~\mathrm{m}^2$. Substitution into Eq.~\ref{eq}, followed by Eqs.~\ref{eq} and \ref{eq}, yields $t_H \approx 827.5 \, \mathrm{fs}$,
corresponding to 8275 simulation time steps for $\Delta t=0.1 \, \mathrm{fs}$. As an independent check, the Fourier transform of the wave packet autocorrelation function shows spectral resolution emerging near 8000 time steps, in good agreement with this estimate.

\subsection{Computation of the instantaneous spatial participation ratio}

As a measure of the instantaneous spatial localization of the wave packet, we compute the spatial participation ratio $\Gamma_S(t)$, which quantifies the degree of localization or delocalization of the coordinate-space probability density over the accessible area:
\begin{equation}
\Gamma_S(t) = A \int!!\int_A |\Psi(x, y, t)|^4, dx,dy,
\label{eq}
\end{equation}
where the integral is taken over the billiard area $A$. In the numerical simulations, this quantity is evaluated as
\begin{equation}
\Gamma_S(t) \approx A \sum_{x, y} |\Psi(x, y, t)|^4, \Delta x, \Delta y.
\label{eq}
\end{equation}

For a spatially uniform probability density, $\Gamma_S(t)$ approaches unity. In the present system, however, chaotic wavefunctions exhibit spatial fluctuations associated with their nodal structure, yielding values around $0.5$ for Berry-like states.

\subsection{Computation of the number of phase space cells accessed}

To quantify the evolution of phase-space ergodicity, we compute the effective number of phase-space cells accessed by the wave packet over a finite time interval $t$. This quantity measures the extent of phase-space exploration and, as derived in the main text, is related to the time-integrated autocorrelation function in the following manner:
\begin{equation}
\begin{split}
\frac{1}{\mathcal{N}t} &= h^D \text{Tr}\left[\bar{\rho}^2\right]\\ 
&= \frac{2}{t} \int_0^t \left(1 - \frac{\tau}{t} \right) P(\tau) \, d\tau \\ &\longrightarrow \frac{2}{t} \sum_{\tau = 0}^t \left(1 - \frac{\tau}{t} \right) P(\tau) \, \Delta \tau,
\end{split}
\end{equation}
where $h$ is Planck’s constant, $D=2$ is the number of degrees of freedom, and $\Delta\tau$ is the timestep. The autocorrelation weighting accounts for the finite observation time and partial occupation of phase-space cells, providing a resolution-dependent measure of the effective phase-space volume explored by the wave packet.

Our simulations show that the number of effectively accessed phase-space cells saturates at approximately the Heisenberg time $t_H$. At longer times, no substantial increase in the number of occupied cells is observed, although the degree of occupation within the already accessed cells may continue to evolve.

\end{document}